\documentclass[12pt]{article}

\font\grande=cmr9.5 scaled \magstep4
\font\medio=cmr9.5 scaled \magstep2
\outer\def\beginsection#1\par{\medbreak\bigskip
      \message{#1}\leftline{\bf#1}\nobreak\medskip
\vskip-\parskip
      \noindent}
\usepackage{graphicx} 
\begin{document}
\bibliographystyle {unsrt}

\titlepage

\begin{flushright}
CERN-PH-TH/2013-300
\end{flushright}

\vspace{10mm}
\begin{center}
{\grande No-hair conjectures, primordial shear}\\
\vspace{10mm}
{\grande and protoinflationary initial conditions}\\
\vspace{1.5cm}
 Massimo Giovannini
 \footnote{Electronic address: massimo.giovannini@cern.ch}\\
\vspace{1cm}
{{\sl Department of Physics, 
Theory Division, CERN, 1211 Geneva 23, Switzerland }}\\
\vspace{0.5cm}
{{\sl INFN, Section of Milan-Bicocca, 20126 Milan, Italy}}
\vspace*{0.5cm}
\end{center}

\vskip 0.5cm
\centerline{\medio  Abstract}
Anisotropic inflationary background geometries are analyzed in the context of an extended gauge action where 
the electric and magnetic susceptibilities are not bound to coincide and depend on the inflaton field. 
After deriving various classes of solutions with electric and  magnetic hairs, we discuss the problem of the initial boundary 
conditions of the shear parameter and consider a globally neutral plasma as a possible relic of a preinflationary stage of expansion.
While electric hairs are washed out by the finite value of the protoinflationary conductivity, magnetic hairs can persist and introduce a tiny amount of shear causing a different inflationary rate of expansion along orthogonal spatial directions. The plasma interactions are a necessary criterion to discriminate between physical and unphysical initial 
conditions but they are not strictly sufficient to warrant the stability of a given magnetic solution. 
\vskip 0.5cm

\noindent

\vspace{5mm}

\vfill
\newpage
\renewcommand{\theequation}{1.\arabic{equation}}
\setcounter{equation}{0}
\section{Introduction}
\label{sec1}
The fate of anisotropies in the expansion of the primeval plasma has been a recurrent theme of discussion 
since the early analyses of Lifshitz, Khalatnikov and Belinskii \cite{lif1,lif2}, 
Hoyle and Narlikar \cite{hoyle}, Zeldovich and collaborators \cite{zel1}, Misner \cite{misner}, Rees \cite{rees} and others. The attention 
has been originally focussed on the possibility of chaotic initial conditions for a standard (and mostly decelerated) stage of expansion.
The dynamics of the shear and of the spatial gradients of the geometry have been subsequently revisited
in accelerated background geometries in connection with the inflationary hypothesis.

One of the basic motivations of the inflationary paradigm (see e.g. \cite{wein}) is to wash out primeval anisotropies in the expansion as soon as the inflationary event horizon is formed (see \cite{barrow} for a lucid account of this perspective). Similar statements and expectations hold for the spatial gradients 
of the geometry that are argued to be exponentially suppressed during a de Sitter or quasi-de Sitter stage of expansion \cite{star,wald}. These 
conclusions can be reached in the framework of the gradient expansion pioneered in Ref. \cite{lif1,lif2} and are at the heart of 
the  {\em cosmic non-hair conjecture} stipulating that in conventional inflationary models any finite portion of the universe gradually loses the memory of an initially imposed anisotropy or inhomogeneity so that the universe attains the observed regularity regardless of the initial boundary conditions \cite{hoyle}. 

The implications of the no-hair conjecture have been questioned long ago by Barrow who showed that for a class of power-law inflationary backgrounds the universe does not need to approach local isotropy and homogeneity \cite{barrow}. A similar perspective was invoked later in the context of bouncing models \cite{mg1} where it was demonstrated that the degree of isotropy depends on the dynamics and on the duration of the bounce. Tiny amounts of anisotropy could be achieved either perturbatively (i.e. by looking at the parametric amplification of quantum fluctuations in anisotropic backgrounds) or non perturbatively (i.e. by breaking the local isotropy of the spatial metric with the inclusion of a background gauge field). A key role is played, in this context, by quadratic curvature corrections to the Einstein-Hilbert action.

It seems odd to concoct inflationary scenarios where the memory of initial conditions is preserved, at least in some form. In spite of this motivated viewpoint, the imprint of a tiny anisotropy in the expansion has been recently revived in connection with the analysis of the temperature and polarisation power spectra of the cosmic microwave background (see, e.g. \cite{wmap1,wmap2,wmap3}). In a region where cosmic variance dominates there have been indications of possible alignments of the lower multipoles of the temperature anisotropies. While it is difficult 
to argue if this is a real physical evidence or rather a systematic effect, various models and discussions appeared in the literature (see \cite{anis1,anis2,anis3} for an incomplete list of references). The aim of these analyses focussed on the possibility of mildly anisotropic inflationary models and can be divided, broadly speaking, into two categories. In some models the anisotropy is realized via higher order corrections to the gravity action and quasi-de Sitter solutions can be found since the differential equations reduce to second order algebraic relations allowing for different expansion rates along orthogonal spatial directions. The other strategy is to couple the gauge kinetic term to some scalar degree of freedom. 

From the earlier discussions of \cite{hoyle,zel1,rees,mg1} an essential ingredient of mildly anisotropic models involve gauge fields. It seems 
therefore interesting to analyze the situation where the electric and magnetic susceptibilities are not bound to coincide during inflation. This possibility has been neglected in the case of anisotropic inflation but received attention as a mechanism for the successful production 
of large-scale magnetic fields \cite{mg2}. For sake of concreteness in what follows we shall be interested in exploring the consequences of the general gauge action: 
\begin{eqnarray}
S_{\mathrm{gauge}} = - \frac{1}{16 \pi} \int \, d^{4} x\, \sqrt{-g} \biggl[ \lambda(\varphi,\psi) F_{\alpha\beta} \, F^{\alpha\beta} + {\mathcal M}_{\sigma}^{\rho}(\varphi,\psi) 
F_{\rho\alpha}\, F^{\sigma\alpha} - {\mathcal N}_{\sigma}^{\rho}(\varphi,\psi) 
\tilde{F}_{\rho\alpha}\, \tilde{F}^{\sigma\alpha} \biggr],
\label{ac1}
\end{eqnarray} 
where $F^{\mu\nu}$ and $\tilde{F}^{\mu\nu}$ are, respectively, the gauge field strength and its dual;
 $g = \mathrm{det}g_{\mu\nu}$ is the determinant of the four-dimensional metric with signature 
 mostly minus. Non-abelian generalisations of Eq. (\ref{ac1}) are possible but shall not be 
 directly discussed here.  The tensors ${\mathcal M}_{\rho}^{\sigma}$ and ${\mathcal N}_{\rho}^{\sigma}$ arise as 
 derivative of scalar degrees of freedom generically denoted, in what follows, by $\varphi$ or $\psi$.
If ${\mathcal N}_{\sigma}^{\rho}=0$ and  ${\mathcal M}_{\rho}^{\sigma} = \partial_{\rho}\psi\partial^{\sigma}\psi$ Eq. (\ref{ac1}) 
appears in the relativistic generalization of Casimir-Polder interactions \cite{such}.  In the absence of ${\mathcal N}_{\sigma}^{\rho}$
 and ${\mathcal M}_{\sigma}^{\rho}$ the action (\ref{ac1}) reduces to the standard case often studied
 in connection with the amplification of large-scale magnetic fields (see \cite{mg2,f2} and references therein).
 Equation (\ref{ac1}) must be complemented by the gravity and scalar actions so that the total action can be 
 symbolically written as:
 \begin{equation}
 S_{\mathrm{tot}} = S_{\mathrm{gravity}}+ S_{\mathrm{scalar}} + S_{\mathrm{gauge}} + S_{\mathrm{fluid}},
\label{actot}
\end{equation}
 where  $S_{\mathrm{gravity}}$, $S_{\mathrm{scalar}}$ and $S_{\mathrm{fluid}}$ denote respectively the gravity, scalar and fluid  contributions. 
 
 To avoid specific tunings of the initial boundary conditions, the inflationary phase 
 is complemented by a protoinflationary epoch where the expansion is not accelerated and
 the matter content is provided by a globally neutral plasma dominated by radiation 
 (see e.g. \cite{nonvac,proto2,proto1} for the analog situation in the isotropic case). 
 The addition of a fluid part in Eq. (\ref{actot}) defines a consistent framework where the problem of the initial conditions can be 
 addressed. The plasma interacts differently with the electric and magnetic hairs: while the former are 
dissipated, the latter are only diluted by the expansion. This criterion pins down solutions that are potentially physical but it is, per se, 
 not sufficient to ensure the dynamical stability of  the corresponding (anisotropic) fixed point.
  
The layout of the paper is the  following. In section \ref{sec2} the equations of motion 
implied by Eq. (\ref{actot}) shall be discussed in covariant and non-covariant terms. In sections \ref{sec3}  and \ref{sec4}
various classes of exact solutions of the system will be derived and  illustrated with particular attention to the role played by the electric and 
magnetic initial conditions. Section \ref{sec5} is devoted to the problem of protoinflationary 
boundary conditions and to the sufficient criteria for the stability of various classes of solutions. Section \ref{sec6} 
contains the concluding remarks.

\renewcommand{\theequation}{2.\arabic{equation}}
\setcounter{equation}{0}
\section{Generalized equations of motion}
\label{sec2}
\subsection{The full set of equations}
The gravity and the scalar parts of the action appearing in Eq. (\ref{actot}) are, respectively, 
\begin{eqnarray}
S_{\mathrm{gravity}} &=& - \frac{1}{16 \pi G} \int \, d^{4}x\, \sqrt{-g} \, R,
\label{gr1}\\
S_{\mathrm{scalar}} &=& \int \, d^{4} x\, \sqrt{-g} \biggl[ \frac{1}{2} g^{\alpha\beta} \partial_{\alpha} \varphi \partial_{\beta} \varphi 
+ \frac{1}{2} g^{\alpha\beta} \partial_{\alpha} \psi \partial_{\beta} \psi - V(\varphi,\psi)\biggr],
\label{scal1}
\end{eqnarray}
where $V(\varphi,\psi)$ denotes the potential depending on the two\footnote{Supplementary scalar degrees of freedom 
 can be included in the discussion and they can coincide with a second inflaton field or with some other spectator field. For illustrative purposes, we shall keep only two fields and even reduce to a single field in sections \ref{sec4} and \ref{sec5}.} scalar fields $\varphi$ and $\psi$. The variation of the total action with respect to the tensor, scalar and vector fields leads to the corresponding equations of motion. In particular the Einstein equations are:
\begin{equation}
{\mathcal G}_{\mu}^{\nu} = 8 \pi G \biggl[ T_{\mu}^{\nu}(\varphi,\psi) + {\mathcal T}_{\mu}^{\nu}(F) + {\mathcal I}_{\mu}^{\nu}(\varphi,\,\psi,\,F) + T_{\mu}^{\nu}(\rho,\,p) \biggr],
\label{EE}
\end{equation}
where the four contributions to the total energy-momentum tensor appearing at the right hand side of Eq. (\ref{EE}) are given, respectively, by:
\begin{eqnarray}
T_{\mu}^{\nu}(\varphi,\psi) &=& \partial_{\mu}\varphi \partial^{\nu} \varphi + \partial_{\mu}\psi \partial^{\nu} \psi - \biggl[ \frac{1}{2} g^{\alpha\beta}\partial_{\alpha} \varphi \partial_{\beta} \varphi + \frac{1}{2} g^{\alpha\beta}\partial_{\alpha} \psi \partial_{\beta} \psi
- V(\varphi,\psi)\biggr],
\label{Tph}\\
{\mathcal T}_{\mu}^{\nu}(F) &=& \frac{1}{4\pi} \biggl[ - {\mathcal S}_{\mu}^{\nu}(F) + \frac{1}{4} {\mathcal S}(F) \delta_{\mu}^{\nu} \biggr],
\label{TF}\\
{\mathcal I}_{\mu}^{\nu}(\varphi,\,\psi,\,F) &=&\frac{1}{8\pi} \biggl[ {\mathcal N}^{\sigma}_{\mu}\, \tilde{F}^{\nu\alpha} \, \tilde{F}_{\sigma\alpha} - {\mathcal M}^{\sigma}_{\mu}\, F^{\nu\alpha} \, F_{\sigma\alpha} \biggr],
\label{INT}\\
T_{\mu}^{\nu}(\rho,\,p) &=& ( p + \rho) u_{\mu} u^{\nu} - p \delta_{\mu}^{\nu}.
\end{eqnarray}
In Eq. (\ref{TF}) the following auxiliary tensor has been introduced:
\begin{eqnarray}
{\mathcal S}_{\mu}^{\nu}(F) = \frac{1}{2} \biggl[ {\mathcal M}^{\rho}_{\mu} F_{\rho\alpha} \, F^{\nu\alpha} + {\mathcal M}^{\rho}_{\sigma} \, F_{\rho\mu} \, F^{\sigma\nu} - {\mathcal N}^{\rho}_{\mu} \, \tilde{F}_{\rho\alpha} \, \tilde{F}^{\nu\alpha} - {\mathcal N}^{\rho}_{\sigma} \, \tilde{F}_{\rho\mu} \, \tilde{F}^{\sigma\nu} + 2 \lambda F_{\alpha\mu}\, F^{\alpha\nu}\biggr];
\label{St}
\end{eqnarray}
${\mathcal T}_{\mu}^{\nu}(F)$ is the energy-momentum tensor of the gauge field while ${\mathcal I}_{\mu}^{\nu}(\varphi, \psi, F)$ 
is the energy-momentum tensor arising from the interaction of the gauge fields with the scalar fields. The equations obeyed by $\varphi$ and $\psi$ are:
\begin{eqnarray}
&& g^{\alpha\beta} \nabla_{\alpha} \nabla_{\beta} \varphi + \frac{\partial V}{\partial \varphi}  + \frac{{\mathcal Q}_{\varphi}}{16 \pi} =0,
\label{pheq}\\
&&  g^{\alpha\beta} \nabla_{\alpha} \nabla_{\beta} \psi + \frac{\partial V}{\partial \psi} + \frac{{\mathcal Q}_{\psi}}{16 \pi} =0,
\label{pseq}
\end{eqnarray}
where $\nabla_{\alpha}$ denotes the covariant derivative; moreover $Q_{\varphi}$ and $Q_{\psi}$ are:
\begin{eqnarray}
{\mathcal Q}_{\varphi} &=& \frac{\partial \lambda}{\partial \varphi} F_{\alpha\beta} \, F^{\alpha\beta} + \frac{\partial {\mathcal M}^{\rho}_{\sigma}}{\partial\varphi} F_{\alpha\beta} F^{\alpha\beta} - \frac{\partial {\mathcal N}^{\rho}_{\sigma}}{\partial\varphi} \tilde{F}_{\alpha\beta} \tilde{F}^{\alpha\beta},
\label{Qphi}\\
{\mathcal Q}_{\psi} &=&\frac{\partial \lambda}{\partial \psi} F_{\alpha\beta} \, F^{\alpha\beta} + \frac{\partial {\mathcal M}^{\rho}_{\sigma}}{\partial\psi}  F_{\alpha\beta} \,F^{\alpha\beta}- \frac{\partial {\mathcal N}^{\rho}_{\sigma}}{\partial\psi} \tilde{F}_{\alpha\beta} \tilde{F}^{\alpha\beta}.
\label{Qpsi}
\end{eqnarray}
Defining the pair of antisymmetric tensors ${\mathcal Z}^{\mu\nu} $ and ${\mathcal W}^{\mu\nu} $:
\begin{eqnarray}
{\mathcal Z}^{\mu\nu} &=&\biggl( {\mathcal M}^{\mu}_{\sigma} \, F^{\sigma\nu}  - {\mathcal M}^{\nu}_{\sigma} \, F^{\sigma\mu}\biggr),
\label{Z}\\
{\mathcal W}^{\mu\nu} &=& \, {\mathcal N}^{\rho}_{\sigma} \, E^{\sigma \alpha \mu\nu}\, \tilde{F}_{\rho\alpha}
= {\mathcal N}^{\nu}_{\,\rho} \, F^{\mu\rho} -  {\mathcal N}^{\mu}_{\,\rho} \, F^{\nu\rho} - {\mathcal N}^{\rho}_{\,\rho} \, F^{\mu\nu}, 
\label{W}
\end{eqnarray}
the equations of the gauge fields are:
\begin{eqnarray}
&& \nabla_{\mu} \biggl( \lambda\, F^{\mu\nu} \biggr) +  \frac{1}{2} \nabla_{\mu} {\mathcal Z}^{\mu\nu} -  \frac{1}{2}\nabla_{\mu} {\mathcal W}^{\mu\nu} = 4\pi j^{\nu},
\label{Feq}\\
&& \nabla_{\mu} \, \tilde{F}^{\mu\nu} =0. 
\label{tFeq}
\end{eqnarray}

\subsection{Covariant decompositions}
The four-dimensional rank-two tensors ${\mathcal M}_{\rho\sigma}$ and ${\mathcal N}_{\rho\sigma}$ can be 
covariantly decomposed as follows\footnote{The decomposition (\ref{dec}) can be applied to ${\mathcal N}_{\rho\sigma}$ but the explicit expressions shall not be repeated.}:
\begin{equation}
{\mathcal M}_{\rho\sigma} = {\mathcal V}\, u_{\rho}\, u_{\sigma} + q_{\rho} u_{\sigma} + p_{\sigma} u_{\rho} + f_{\rho\sigma},
\label{dec}
\end{equation}
where $g^{\alpha\beta} \, u_{\alpha} \, u_{\beta} =1$, ${\mathcal M}_{\rho\sigma} u^{\rho} u^{\sigma} = {\mathcal V}$ and $q_{\alpha} u^{\alpha} = p_{\beta} u^{\beta} = u^{\alpha} f_{\alpha\beta} =0$.  Introducing the projector $h^{\alpha}_{\mu} = \delta_{\mu}^{\alpha} - u_{\mu} \, u^{\alpha}$, 
the last term at the right hand side of Eq. (\ref{dec}) (i.e. $f_{\rho\sigma}$) can be further 
separated into a symmetric part (containing the trace-full and the trace-free contributions) supplemented by an antisymmetric part. 
At the end of this straightforward procedure ${\mathcal M}_{\rho\sigma}$ reads:
\begin{equation}
{\mathcal M}_{\rho\sigma} = {\mathcal V}\, u_{\rho}\, u_{\sigma} + q_{\rho} u_{\sigma} + p_{\sigma} u_{\rho} + \frac{1}{3} {\mathcal M}_{\mu\nu} \, h^{\mu\nu} \, h_{\rho\sigma} + \overline{{\mathcal M}}_{\rho_\sigma} + {\mathcal M}_{[\mu\,\nu]} h^{\mu}_{\rho} \, h^{\nu}_{\sigma},
\label{dec2}
\end{equation}
where $\overline{{\mathcal M}}_{\rho\sigma} = h^{\mu}_{\rho}\, h^{\nu}_{\sigma} [ {\mathcal M}_{(\mu\,\nu)} - {\mathcal M}_{\alpha\beta}h^{\alpha\beta}\, h_{\rho\sigma}/3]$ 
is the trace-free contribution; $ {\mathcal M}_{(\mu\,\nu)}$ and $ {\mathcal M}_{[\mu\,\nu]}$ denote, as usual, 
the symmetric and the antisymmetric parts of the corresponding tensor.  

If ${\mathcal M}_{\rho}^{\sigma}$ and ${\mathcal N}_{\rho}^{\sigma}$ arise as derivatives of a scalar degree of freedom they can be expressed as\footnote{If ${\mathcal M}_{\rho}^{\sigma}$ and ${\mathcal N}_{\rho}^{\sigma}$ are both vanishing in Eq. (\ref{actot}) 
the magnetic and the electric susceptibilities are coincident. If either ${\mathcal M}_{\rho}^{\sigma}$ or ${\mathcal N}_{\rho}^{\sigma}$ are different 
from zero then the electric and the magnetic susceptibilities will be different, as recently discussed in a different context \cite{mg2}. 
If either ${\mathcal M}_{\rho}^{\sigma}$ or ${\mathcal N}_{\rho}^{\sigma}$ are proportional to $\delta_{\rho}^{\sigma}$,
$\lambda(\varphi,\psi)$  is redefined. For instance, if
${\mathcal N}_{\rho}^{\sigma} = A(\varphi,\,\psi) \, \delta_{\rho}^{\sigma}$ the new effective coupling multiplying the gauge kinetic 
term becomes $\lambda(\varphi,\,\psi) + A(\varphi,\,\psi)$. } 
\begin{equation}
{\mathcal M}_{\rho}^{\sigma}(\varphi) = \lambda_{E}(\varphi)\, U_{\rho}\, U^{\sigma}, \qquad {\mathcal N}_{\rho}^{\sigma}(\psi) = \lambda_{B}(\psi)\, \overline{U}_{\rho}\, \overline{U}^{\sigma},
\label{dec3}
\end{equation}
where 
\begin{equation}
U^{\rho} = \frac{\partial^{\rho} \varphi}{\sqrt{g^{\alpha\beta} \partial_{\alpha} \varphi\partial_{\beta} \varphi}}, \qquad \overline{U}^{\rho} = \frac{\partial^{\rho} \psi}{\sqrt{g^{\alpha\beta} \partial_{\alpha} \psi\partial_{\beta} \psi}},
\label{dec4}
\end{equation}
so that $g_{\alpha\beta} \, U^{\alpha} U^{\beta} =1$ and $g_{\alpha\beta} \, \overline{U}^{\alpha} \overline{U}^{\beta} =1$.
The functionals $\lambda_{E}(\varphi)$ and $\lambda_{B}(\psi)$ are naturally associated with the electric and the magnetic degrees of freedom.
In fact $u_{\rho} \tilde{F}^{\alpha\rho} = {\mathcal B}^{\alpha}$ and $u_{\rho} F^{\alpha\rho} = {\mathcal E}^{\alpha}$ are the electric and magnetic fields in covariant form as it follows from the generally covariant decomposition of the gauge field strengths \cite{lic}:
\begin{eqnarray}
F_{\alpha\beta} &=& {\mathcal E}_{\alpha} u_{\beta} - {\mathcal E}_{\beta} u_{\alpha} + E_{\alpha\beta\rho\sigma} \, u^{\rho} \, {\mathcal B}^{\sigma},
\nonumber\\
\tilde{F}^{\alpha\beta} &=& {\mathcal B}^{\alpha} u^{\beta} - {\mathcal B}^{\beta} u^{\alpha} + E^{\alpha\beta\rho\sigma} \, {\mathcal E}_{\rho} \, u_{\sigma},
\label{COVDEC}
\end{eqnarray}
where $E_{\alpha\beta\rho\sigma} = \sqrt{-g}\, \epsilon_{\alpha\beta\rho\sigma}$ and  $\epsilon_{\alpha\beta\rho\sigma}$ is the Levi-Civita symbol in 4 dimensions. 
To make contact with different notations employed in the literature let us remark that the shear tensor
is customarily defined as 
\begin{equation}
\sigma_{\alpha\beta} = h^{\mu}_{\alpha} \, h^{\nu}_{\beta} \, \biggl[ \frac{1}{2} \biggl( \nabla_{\mu} u_{\nu} + \nabla_{\nu} u_{\mu}\biggr) - 
\frac{1}{3} \nabla_{\gamma} u_{\delta} \, h^{\gamma\delta} h_{\mu\nu} \biggr],
\label{shear}
\end{equation} 
measuring the difference in the expansion along the different directions.  Instead of dealing with $\sigma_{\alpha\beta}$ we shall deal 
preferentially with the so-called shear parameter \cite{zel1} given by the ratio between the shear tensor and the 
mean expansion rate.

\subsection{ADM decomposition}
So far the discussion has been conducted in covariant language\footnote{The Greek indices will run over the 
four space-time dimensions while the Latin indices will denote the spatial 
indices.} but for the forthcoming applications
the metric tensor can be decomposed as 
\begin{equation}
g_{00} = N^2 - N_{k} N^{k},\qquad g_{ij} = - \gamma_{ij},\qquad g_{0i} = - N_{i},
\label{ADM1}
\end{equation}
where  $N$, $N^{i}$ and $\gamma_{ij}$ 
denote, respectively, the lapse function, the shift vector and the three-dimensional metric.
According to the Arnowitt, Deser and Misner (ADM) decomposition  \cite{ADM1} of Eq. (\ref{ADM1}) , 
the extrinsic curvature 
$K_{ij}$ and the spatial components of the Ricci tensor $r_{ij}$ become:
\begin{eqnarray}
K_{ij} &=& \frac{1}{2 N} \biggl[- \partial_{\tau}\gamma_{ij} + ^{(3)}\nabla_{i}N_{j} + ^{(3)}\nabla_{j} N_{i} 
\biggr],
\label{ADM1a}\\
{\mathcal R}_{ij} &=& \partial_{m} \, ^{(3)}\Gamma^{m}_{ij} -\partial_{j} ^{(3)}\Gamma_{i m}^{m} + ^{(3)}\Gamma_{i j}^{m} 
\,^{(3)}\Gamma_{m n}^{n} - ^{(3)}\Gamma_{j n}^{m} \,^{(3)}\Gamma_{i m}^{n},
\label{ADM1b}
\end{eqnarray}
where  $^{(3)}\nabla_{i}$ is the covariant derivative defined 
with respect to the metric $\gamma_{ij}$, $\partial_{\tau}$ denotes a derivation with respect to the time coordinate 
$\tau$ and $^{(3)}\Gamma_{i j}^{m}$ are the Christoffel symbols computed from $\gamma_{ij}$.
The contracted form of Eq. (\ref{EE}) is 
\begin{equation}
R_{\mu}^{\nu} = \ell_{\mathrm{P}}^2\, \biggl[ \partial_{\mu} \varphi \partial^{\nu} \varphi + \partial_{\mu} \psi \partial^{\nu} \psi + {\mathcal T}_{\mu}^{\nu} + (p + \rho) u_{\mu} u^{\nu} + {\mathcal I}_{\mu}^{\nu}
+  \frac{1}{2}\biggl( p- \rho  - 2 V(\varphi,\psi)   - {\mathcal I}\biggr)\delta_{\mu}^{\nu}\biggr],
\label{EE1}
\end{equation}
where $\ell_{\mathrm{P}}=\sqrt{8 \pi G}$.  Using Eqs.(\ref{ADM1}), (\ref{ADM1a}) and (\ref{ADM1b}), the $(00)$, $(0i)$ and $(ij)$ components 
of Eq. (\ref{EE1}) are:
\begin{eqnarray}
&& \partial_{\tau} K - N \,\mathrm{Tr} K^2 + \nabla^2 N = N\, \ell_{\mathrm{P}}^2 \, {\mathcal P}_{0}^{0},
\label{EE3}\\
&& \nabla_{i} K - \nabla_{k} K^{k}_{i} = N \,\ell_{\mathrm{P}}^2 \, {\mathcal P}_{i}^{0}, 
\label{EE4}\\
&& \partial_{\tau} K_{i}^{j} - N\, K\, K_{i}^{j} - N\, {\mathcal R}_{i}^{j} + \nabla_{i} \nabla^{j} N = N\, \ell_{\mathrm{P}}^2\, {\mathcal P}_{i}^{j},
\label{EE5}
\end{eqnarray}
where the source terms ${\mathcal P}_{0}^{0}$,  ${\mathcal P}_{i}^{0}$ and ${\mathcal P}_{i}^{j}$ can be expressed as\footnote{For sake  of simplicity,  the shorthand notation $u^2 = \gamma^{ij} u_{i} u_{j}$ has been adopted.}:
\begin{eqnarray}
{\mathcal P}_{0}^{0} &=& \frac{3 p + \rho}{2} 
 + ( p + \rho) \, u^2 + {\mathcal T}_{0}^{0} + \biggl({\mathcal I}_{0}^{0} - \frac{{\mathcal I}}{2} \biggr) 
 \nonumber\\
 &+& \frac{{\dot{\varphi}}^2 + {\dot{\psi}}^2}{N^2} - \gamma^{ij} \partial_{i} \varphi \partial_{j} \varphi - \gamma^{ij} \partial_{i} \psi \partial_{j} \psi - V
 \label{P00}\\
 {\mathcal P}_{i}^{0} &=&  \frac{u_{i}}{N} ( p + \rho) \sqrt{1 + u^2} + {\mathcal T}_{i}^{0} + {\mathcal I}_{i}^{0}  + \frac{\partial_{i} \varphi \dot{\varphi}}{N^2} + 
 \frac{\partial_{i} \psi \dot{\psi}}{N^2},
 \label{Pi0}\\
 {\mathcal P}_{i}^{j} &=&  \frac{p - \rho}{2} \delta_{i}^{j} - ( p + \rho) u_{i} u^{j} +  {\mathcal T}_{i}^{j} + \biggl({\mathcal I}_{i}^{j} - \frac{{\mathcal I}}{2} \delta_{i}^{j}\biggr) 
 \nonumber\\
 &-& \partial_{i} \varphi \partial^{j} \varphi - \partial_{i}\psi \partial^{j} \psi - V\delta_{i}^{j}.
\label{Pij}
\end{eqnarray}
In what follows the derivation with respect to the time coordinate will be denoted by the overdot.
The system of Eqs. (\ref{EE3}), (\ref{EE4}) and (\ref{EE5}) must be supplemented by the explicit 
form of Eqs. (\ref{pheq})--(\ref{pseq}) and of the covariant conservation of the fluid energy-momentum tensor. 
These formulas shall not be reported here but  their explicit expressions shall be given directly in the specific cases 
discussed hereunder.

\subsection{Electric and magnetic hairs}
Even if different Bianchi classes can be discussed with similar methods (see e.g. \cite{mg0}) we shall focus the attention on the Bianchi type-I geometries so that, according to Eq. (\ref{ADM1}),
\begin{equation}
N=1, \qquad \gamma_{x\,x}(t)= a^2(t), \qquad \gamma_{y\,y}(t) = \gamma_{z\, z}(t) = b^2(t).
\label{FF0}
\end{equation}
Equation (\ref{FF0}) implies that 
Eqs. (\ref{Feq}), (\ref{tFeq}), (\ref{Z}) and (\ref{W}) can support either magnetic or electric initial conditions.
More specifically, in the metric (\ref{FF0}) the gauge field equations are solved by the following configuration: 
\begin{eqnarray}
F_{yz} &=& - B, \qquad F^{0\,x} = - \frac{E(t)}{ N \sqrt{\gamma}} = - \frac{E(t)}{ a\, b^2}, 
\nonumber\\
\tilde{F}^{0\,x} &=& - \frac{B}{N \, \sqrt{\gamma}} = - \frac{B}{a b^2}.
\label{FF1}
\end{eqnarray}
From the purely geometric point viewpoint both solutions respect the symmetry of the geometry and are therefore plausible. There however are important physical differences between the two.
Since the initial conditions of conventional inflationary models are not set in the vacuum but rather 
during a protoinflationary phase where the universe is globally neutral, the electric and magnetic degrees of freedom 
have to undergo a different evolution (see section \ref{sec5}). Thus in the absence of Ohmic currents, $E(t)$ is constant. Conversely, if Ohmic currents are included in Eq. (\ref{Feq}) as 
$j^{\nu} = \sigma \, F^{\mu\,\nu}\, u_{\mu}$  the equation obeyed by $E(t)$ becomes:
\begin{equation}
\dot{E} + 4 \pi \frac{\sigma}{\lambda} E =0,
\label{decay}
\end{equation}
where sigma denotes the conductivity and must not be confused with the shear tensor $\sigma_{\alpha\beta}$ 
defined earlier in Eq. (\ref{shear}). Note, finally, that since $N=1$ the derivation with respect to $\tau$ coincides 
with the derivation with respect to $t$ (i.e. the cosmic time coordinate) that is denoted throughout the paper by an overdot.
\renewcommand{\theequation}{3.\arabic{equation}}
\setcounter{equation}{0}
\section{Anisotropic inflationary solutions}
\label{sec3}
\subsection{Equal susceptibilities}
Let us consider the situation where $V= V(\varphi)$ and ${\mathcal M}^{\rho}_{\sigma} = {\mathcal N}^{\rho}_{\sigma} =0$.  Equations (\ref{EE3}), (\ref{EE4}) and (\ref{EE5}) can be made explicit with the help of Eqs. (\ref{FF0})--(\ref{FF1}) and they become\footnote{The spatial gradients will be neglected hereunder since we are concerned with 
anisotropic but fully homogeneous solutions. }:
\begin{eqnarray}
&& \dot{H} + 2 \dot{F} + ( H^2 + 2 F^2) = - \frac{1}{\overline{M}_{\mathrm{P}}^2} \biggl[ \dot{\varphi}^2 - V + \frac{ 3 p +\rho}{2} + \frac{\rho_{E}}{\lambda} + 
\lambda\, \rho_{B} \biggr], 
\label{00c1}\\
&& \dot{H} + H ( H + 2 F) = - \frac{1}{\overline{M}_{\mathrm{P}}^2} \biggl[ \frac{p - \rho}{2} - V + \frac{\rho_{E}}{\lambda} + \lambda\, \rho_{B} \biggr],
\label{xxc1}\\
&& \dot{F} + F ( H + 2 F) =  - \frac{1}{\overline{M}_{\mathrm{P}}^2} \biggl[ \frac{p - \rho}{2} - V - \frac{\rho_{E}}{\lambda} - \lambda\, \rho_{B} \biggr],
\label{yyc1}
\end{eqnarray}
where $\overline{M}_{\mathrm{P}} = 1/\ell_{\mathrm{P}}$. The equations for $\varphi$ and for the fluid component are:
\begin{eqnarray}
&& \ddot{\varphi} + ( H + 2 F) \dot{\varphi} + \frac{\partial V}{\partial \varphi} + \frac{1}{\lambda} \frac{\partial \lambda}{\partial \varphi} \biggl( \lambda\, \rho_{B} - \frac{\rho_{E}}{\lambda}\biggr) =0,
\label{phi1}\\
&& \dot{\rho} + ( H + 2 F) (\rho + p)=0,
\label{rho1}
\end{eqnarray}
where $\rho_{B}$ and $\rho_{E}$ are the magnetic and the electric energy densities defined, respectively, as $\rho_{B} = B^2/(8\pi b^4)$ and $\rho_{E} = E^2/(8\pi b^4)$. Summing up term by term Eqs. (\ref{00c1}), (\ref{rho1}) and twice Eq. (\ref{yyc1}) we obtain:
\begin{equation}
2 (\dot{H} + 2 \dot{F}) + ( H^2 + 2 F^2) + ( H+ 2 F)^2 =   \frac{1}{\overline{M}_{\mathrm{P}}^2} \biggl( 4 V - \dot{\varphi}^2 + \rho - 3 p\biggr). 
\label{ham1}
\end{equation}
In the case of Eq. (\ref{FF0}) the components shear tensor
of Eq. (\ref{shear}) are $\sigma_{x}^{x} =  2 ( H- F)/3$ and $\sigma_{y}^{y}= \sigma_{z}^{z} =(F - H)/3$. Subtracting Eq. (\ref{yyc1}) from Eq. (\ref{xxc1}) the evolution of $(H-F)$ can be readily obtained 
\begin{equation}
\dot{H} - \dot{F}  + (H - F) (H+ 2 F) = - \frac{2}{\overline{M}_{\mathrm{P}}^2} \biggl( \frac{\rho_{E}}{\lambda} + \lambda \,\rho_{B} \biggr).
\label{sh2}
\end{equation}

In view of the forthcoming applications, it is practical to rephrase Eqs. (\ref{00c1}), (\ref{xxc1}) and (\ref{yyc1}) 
in terms of the following pair of  Zeldovich variables \cite{zel1}:
\begin{equation}
n = \frac{H + 2 F}{3}, \qquad r = \frac{H - F}{ n},  
\label{zelvar}
\end{equation}
measuring, respectively, the mean expansion rate an the normalized shear parameter.
The anisotropic Hubble rates $H$ and $F$, expressed in terms of $n$ and $r$, can be inserted into Eqs. (\ref{ham1})--(\ref{sh2}) and the following two equations for $n$ and $r$ can be easily derived\footnote{Equations (\ref{n1}) and (\ref{r1}) are suitable for the analysis the approximate solutions holding in the limit $r < 1$, as we shall see. }:
\begin{eqnarray}
&& \dot{n} + 2 n^2 \biggl( 1 + \frac{r^2}{18}\biggr) = \frac{1}{6 \overline{M}_{\mathrm{P}}^2} \biggl[ \rho + \rho_{\varphi} - 3 ( p + p_{\varphi}) \biggr],
\label{n1}\\
&& \dot{r} + n r \biggl( 1 - \frac{r^2}{9} \biggr) = - \frac{r }{6 \, n\,\overline{M}_{\mathrm{P}}^2} (\rho_{\varphi} - 3 p_{\varphi}) - \frac{2}{\overline{M}_{\mathrm{P}}^2\, n} \biggl(\frac{\rho_{E}}{\lambda}  + \lambda\,\rho_{B}\biggr),
\label{r1}
\end{eqnarray}
where $\rho_{\varphi}$ and $p_{\varphi}$ denote the effective energy density and pressure of $\varphi$: 
\begin{equation}
\rho_{\varphi} = \frac{\dot{\varphi}^2}{2} + V(\varphi), \qquad p_{\varphi} = \frac{\dot{\varphi}^2}{2} - V(\varphi).
\end{equation}
In the limits  $\lambda \to 1$ and $\rho_{\varphi} = p_{\varphi}=0$ in Eq. (\ref{r1}) the standard evolution 
of the shear parameter in a decelerated background geometry can be obtained\footnote{Neglecting $r^2$ 
terms, the shear parameter is solely determined, in this case, by the gauge field and the known results on the evolution 
of $r$ in decelerated background geometries can be derived (see e.g. \cite{zel1,mg1,mg0}). See also the discussion in section \ref{sec5}.}. 
The equations obeyed by $\rho_{\varphi}$ and $\rho$ are 
\begin{eqnarray} 
&& \dot{\rho}_{\varphi} + 3 n (\rho_{\varphi} + p_{\varphi}) + \frac{\dot{\lambda}}{\lambda} \biggl(\lambda \rho_{B} - \frac{\rho_{E}}{\lambda} \biggr) =0,
\label{rph}\\
&&\dot{\rho} + 3 n ( \rho + p) =0.
\label{rrf}
\end{eqnarray}
Equation (\ref{rph}) is equivalent to Eq. (\ref{phi1}). 

\subsection{Electric and magnetic solutions}
Setting then to zero  the fluid sources (i.e. $\rho = p =0$), Eqs. (\ref{00c1})--(\ref{phi1}) can be solved by assuming a power-law form for the scale factors:
\begin{equation}
a(t) = \biggl(\frac{t}{t_{*}}\biggr)^{\alpha}, \qquad b(t) = \biggl(\frac{t}{t_{*}}\biggr)^{\beta},\qquad \dot{\varphi} = \frac{\varphi_{1} \, \overline{M}_{\mathrm{P}}}{t}.
\label{ans1}
\end{equation}
It is practical to solve the whole system by separating the various contributions by means of appropriate linear combinations.
More specifically, from the sum of Eqs. (\ref{xxc1}) and (\ref{yyc1}) the resulting equation only contains the scalar potential:
\begin{equation}
\dot{H} + \dot{F} + ( H + F) ( H + 2 F) = \frac{2 V}{\overline{M}_{\mathrm{P}}^2}.
\label{con1}
\end{equation}
Inserting Eq. (\ref{ans1}) into Eq.(\ref{con1})  the first condition to be 
satisfied by the actual solution is:
\begin{equation}
(\alpha + \beta) ( \alpha + 2 \beta -1) = \frac{2 \, t^2\, V}{\overline{M}_{\mathrm{P}}^2}.
\label{con1a}
\end{equation}
With the same logic, subtracting Eq. (\ref{xxc1}) from Eq. (\ref{00c1}) we obtain an equation containing only the kinetic energy of the inflaton:
\begin{equation}
\dot{F} + F ( F - H) = - \frac{{\dot{\varphi}}^2}{2 \, \overline{M}_{\mathrm{P}}^2}.
\label{con2}
\end{equation}
Inserting then Eq. (\ref{ans1}) into Eq. (\ref{con2}) the second condition to be satisfied by $\alpha$ and $\beta$ is
\begin{equation}
\beta( \beta - \alpha -1) = - \frac{\dot{\varphi}^2 t^2}{2 \overline{M}_{\mathrm{P}}^2}. 
\label{con2a}
\end{equation}
The difference between Eqs. (\ref{yyc1}) and (\ref{xxc1}) leads to a relation that involves only the magnetic and the 
electric energy densities:
\begin{equation}
\dot{H} - \dot{F} + (H - F) (H + 2 F) = - \frac{2}{\overline{M}_{\mathrm{P}}^2} \biggl( \frac{\rho_{E}}{\lambda} + \lambda \rho_{B} \biggr).
\label{con3}
\end{equation}
This time the explicit form of Eq. (\ref{con3}) in terms of the parametrization of Eq. (\ref{ans1}) is:
\begin{equation}
(\alpha - \beta) ( \alpha + 2 \beta -1) = -\,\frac{2 t^2}{\overline{M}_{\mathrm{P}}^2} \biggl( \frac{\rho_{E}}{\lambda} + \lambda \rho_{\mathrm{B}}\biggr).
\label{con3a}
\end{equation}
Finally inserting Eq. (\ref{ans1}) into Eq. (\ref{phi1}) the following equation can be obtained:
\begin{equation}
\varphi_{1} ( \alpha + 2 \beta -1) + \frac{t^2}{\overline{M}_{\mathrm{P}}} \biggl(\frac{\partial V}{\partial\varphi} \biggr) + 
\frac{t^2}{\overline{M}_{\mathrm{P}}} \frac{1}{\lambda} \biggl(\frac{\partial \lambda}{\partial\varphi} \biggr)\biggl(\lambda \rho_{B} - \frac{\rho_{E}}{\lambda} \biggr) =0.
\label{con4a}
\end{equation}

Equations (\ref{con1a}), (\ref{con2a}) and (\ref{con4a}) fix consistently $\alpha$ and $\beta$ if the potential  and the susceptibility are exponentials of the inflaton $\varphi$, i.e.\footnote{Note that, incidentally, the $\gamma$ appearing in Eq. (\ref{ans1a}) cannot be confused with the determinant of the spatial part of the metric introduced in the 
previous section since these two quantities will never appear simultaneously in the discussion.}
\begin{equation}
V(\varphi) = V_{0} \exp{( \gamma \,\varphi/\overline{M}_{\mathrm{P}})},\qquad \lambda(\varphi) = \lambda_{0} \exp{( \delta \,\varphi/\overline{M}_{\mathrm{P}})}.
\label{ans1a}
\end{equation}
With these caveats and with the help of Eq. (\ref{ans1a}), Eqs. (\ref{con1a}) and (\ref{con2a}) can be rewritten as:
\begin{equation}
\varphi_{1}^2 = 2 \beta( \alpha + 1 - \beta) \geq 0, \qquad \frac{2 V_{0}}{H_{*}^2\,\overline{M}_{\mathrm{P}}^2} = (\alpha + \beta) ( \alpha + 2 \beta -1) \geq 0, \qquad \gamma \, \varphi_{1} = -\,2,
\label{con6}
\end{equation}
while Eq. (\ref{con4a}) becomes:
\begin{equation}
\varphi_{1}(\alpha + 2 \beta -1) + \frac{\gamma V_{0}}{H_{*}^2 \,\overline{M}_{\mathrm{P}}^2} + \frac{t^2 \, \delta}{\overline{M}_{\mathrm{P}}^2} \biggl( 
\lambda \rho_{B} - \frac{\rho_{E}}{\lambda} \biggr) =0.
\label{con7}
\end{equation}
If $\rho_{E} = 0$ then $\delta = \delta_{B}$; in the opposite case (i.e. $\rho_{B}=0$) $\delta = \delta_{E}$.

In the absence of Ohmic currents $E$ and $B$ are constant in time (see Eq. (\ref{FF1})). 
If $E =0$ and $B \neq 0$ we shall have the {\em magnetic 
solutions}; conversely if $E\neq 0$ and $B=0$ we have the {\em electric solutions}. 
The consistency with Eq. (\ref{con3a}) implies, in the case of magnetic solutions, that
\begin{equation}
\Omega_{B} = \frac{ (\beta - \alpha) ( \alpha + 2\beta -1)}{ 2 \lambda_{0}}, \qquad \delta_{B} = \frac{4 \beta - 2}{ \varphi_{1}}, \qquad 
\Omega_{B} = \frac{B^2}{8 \pi H_{*}^2 \, \overline{M}_{\mathrm{P}}},
\label{con8}
\end{equation}
and $H_{*} = 1/t_{*}$. With the same logic, the consistency of Eq. (\ref{con3a}) with the electric solution implies\footnote{Note that  in Eq. (\ref{ans1a})  $\delta = \delta_{B}$ in the case of magnetic solutions and  $\delta = \delta_{E}$ in the case of electric solutions.}
\begin{equation}
\Omega_{E} = \frac{\lambda_{0} (\beta - \alpha) ( \alpha + 2\beta -1)}{2}, \qquad \delta_{E} = \frac{2 - 4 \beta }{ \varphi_{1}}, \qquad 
\Omega_{E} = \frac{E^2}{8 \pi H_{*}^2}.
\label{con9}
\end{equation}

It is relevant to stress that when $\dot{H}= \dot{F} =0$ the system cannot be reduced to quadratures. The relevant equations are, in this case:
\begin{eqnarray}
&& ( H + F) ( H + 2 F) = 2 \frac{V_{0}}{\overline{M}_{\mathrm{P}}^2}, \qquad (H - F) F = \frac{{\dot{\varphi}}^2}{2 \overline{M}_{\mathrm{P}}^2},
\nonumber\\
&& ( H - F)  (H + 2 F) = - \frac{2}{\overline{M}_{\mathrm{P}}^2} \biggl( \frac{\rho_{E}}{\lambda} + \lambda \rho_{\mathrm{B}}\biggr).
\label{con10}
\end{eqnarray} 
If we ought to have an expanding background with positive scalar kinetic term we must require $H > F$ and $(H+ 2 F)>0$; but these 
two conditions imply that the magnetic and electric energy densities must be negative semidefinite. 
The negative conclusion of Eq. (\ref{con10}) can be evaded if the electric and the magnetic susceptibilities do not coincide, as it will be shown in a class of solutions derived in section \ref{sec4}.

\subsection{Physical constraints on the solutions}
The requirements imposed on $\alpha$ and $\beta$ by the equations of motion can be summarized, in short, 
as follows. The positivity of the potential implies $(\alpha + \beta)(\alpha + 2 \beta -1) \geq 0$.
The positivity of the kinetic term of the inflaton demands
 $\beta(\alpha - \beta +1) \geq 0$; the positivity  of the electromagnetic energy density requires $(\beta - \alpha) ( \alpha + 2 \beta -1) \geq 0$. Focussing on the case $\alpha >0$ and $\beta>0$ we have that the three previous inequalities demand $\beta -1 < \alpha < \beta$, or, which is the same, $\beta = \alpha + \theta$ where $\theta < 1$. This is the fine-tuning we must be prepared to accept if  we ought to have a finite amount of shear. 

So far the electric and magnetic solutions have been parametrized in the $(\alpha,\beta)$ plane but they 
can also be analyzed in the $(\gamma,\delta)$ plane.  For the magnetic initial conditions, Eqs. (\ref{con7}) and (\ref{con8}) give the relation between the two parametrizations: 
\begin{equation}
\alpha = \frac{4}{\gamma (\gamma - \delta_{B})} - \frac{\gamma + \delta_{B}}{2 \gamma}, \qquad \beta= \frac{\gamma - \delta_{B}}{2 \gamma},
\label{conn1}
\end{equation}
where  Eq. (\ref{con6}) has been used  insofar as $\gamma = -2/\varphi_{1}$. 
In the light of Eq. (\ref{conn1}) the condition $(\beta - \alpha) \ll 1$ 
demands that  $\gamma (\gamma - \delta_{B}) > 4$ which also implies $\gamma\ll 1$ while $\delta_{B} \ll -1$. 
Similar conclusions can be derived in the case of electric initial conditions from Eqs. (\ref{con7}) and (\ref{con9}) 
with the result that Eq. (\ref{conn1}) is still valid but with $\delta_{B} \to - \delta_{E}$. The condition $(\beta - \alpha) \ll 1$ reads, for electric solutions, $\gamma(\gamma + \delta_{E}) > 4$ implying $\gamma \ll 1$ and $\delta_{E} \gg 1$. 

Neither the $(\alpha,\beta)$ parametrization nor the $(\gamma,\delta)$ plane are particularly revealing. The most physical parametrization is, in our opinion, a combination of the  Zeldovich variables. More specifically the parameter space of the solution is adequately described by the slow-roll 
parameter $\epsilon$ and by the shear parameter expressed in units of $\epsilon$, i.e. the ratio $ r/\epsilon$.
In terms of $n$ introduced in Eqs. (\ref{zelvar}) and (\ref{n1})--(\ref{r1}), the slow-roll parameter is defined as $\epsilon = - \dot{n}/n^2$. For the standard power-law solutions of this section, each quantity can be 
expressed in terms of $\epsilon$ and in terms of  $\zeta = - r/\epsilon$. Since $\alpha = [ 1/\epsilon +2\, r/(3 \epsilon)]$ and $\beta= [1/\epsilon - r/(3 \epsilon)]$ the condition $(\beta - \alpha) \ll 1$ implies that  $\zeta \ll 1$. For illustration we can write, using Eq. (\ref{con6}),
\begin{equation}
\frac{\dot{\varphi}}{n\, \overline{M}_{\mathrm{P}}} = - \sqrt{6 \epsilon} \sqrt{( 1 + \zeta) \biggl( 1 - \frac{\zeta\, \epsilon}{3}\biggr)} \simeq - \sqrt{6 \epsilon}\,\,\biggl[ 1 + {\mathcal O}(\epsilon) + {\mathcal O}(\zeta)\biggr], 
\label{conn2}
\end{equation}
where the result at the right hand side of the second equality follows in the limit $\epsilon \ll 1$ and $\zeta\ll 1$. 
All the other quantities (i.e. $\gamma$, $\delta_{E}$, $\delta_{B}$ and so on and so forth) can be easily expressed in terms of $\epsilon$ and $\zeta$ and subsequently expanded in powers of $\epsilon$ and $\zeta$, if needed.

\renewcommand{\theequation}{4.\arabic{equation}}
\setcounter{equation}{0}
\section{Anisotropic inflation with different susceptibilities}
\label{sec4}
\subsection{Specific forms of the equations}
In section (\ref{sec3}) we assumed $\lambda_{E} \to 0$ and $\lambda_{B}\to 0$. Some of the solutions obtainable when $\lambda_{E} \neq 0$ and $\lambda_{B} \neq 0$ will now be illustrated without the ambition of being comprehensive\footnote{We shall focus here on the simplest situation, i.e. 
$\lambda_{E}(\varphi)$ and $\lambda_{B}(\varphi)$. There exist also two-field solutions but they are not essential for our discussion and therefore they will not be 
reported here to avoid digressions.}. 
The components of ${\mathcal T}_{\mu}^{\nu}$ (see Eqs. (\ref{EE}) and (\ref{TF})) are:
 \begin{eqnarray}
 {\mathcal T}_{0}^{0} &=& {\mathcal T}_{x}^{x} = - \frac{\lambda + \lambda_{E}/2}{8 \pi} F_{0 x} F^{0 x}  + \frac{\lambda - \lambda_{B}/2}{8\pi } F_{y z} F^{y z}, 
 \nonumber\\
 {\mathcal T}_{y}^{y} &=& {\mathcal T}_{z}^{z} =\frac{\lambda + \lambda_{E}/2}{8 \pi} F_{0 x} F^{0 x} - \frac{\lambda - \lambda_{B}/2}{8\pi } F_{y z} F^{y z}.
\label{Tneq}
 \end{eqnarray}
Moreover, thanks to Eqs. (\ref{FF0})--(\ref{FF1}),  Eq. (\ref{INT}) implies
 \begin{equation}
 {\mathcal I}_{0}^{0} = - \frac{\lambda_{B}}{8\pi} F_{y z} F^{y z}  -  \frac{\lambda_{E}}{8\pi} F_{0 x} F^{0 x}.
\label{I00}
\end{equation}
Equations (\ref{Tneq}) and (\ref{I00}) can be used to obtain the analogs of Eqs. (\ref{00c1})--(\ref{yyc1}) and (\ref{phi1})--(\ref{rho1}). Their explicit forms are:
 \begin{eqnarray}
&& \dot{H} + 2 \dot{F} + ( H^2 + 2 F^2) = - \frac{1}{\overline{M}_{\mathrm{P}}^2} \biggl[ \dot{\varphi}^2 + \frac{ 3 p + \rho}{2} - V +  A_{0} \biggr], 
\label{00neq1}\\
&& \dot{H} + H ( H + 2 F) = - \frac{1}{\overline{M}_{\mathrm{P}}^2} \biggl[ \frac{p - \rho}{2} - V + A_{\parallel} \biggr],
\label{xxneq1}\\
&& \dot{F} + F ( H + 2 F) =  - \frac{1}{\overline{M}_{\mathrm{P}}^2} \biggl[ \frac{p - \rho}{2} - V + A_{\perp}\biggr],
\label{yyneq1}\\
&& \ddot{\varphi} + ( H + 2 F) \dot{\varphi} + \frac{\partial V}{\partial \varphi} + \frac{1}{8\pi} \frac{\partial (\lambda + \lambda_{E}/2)}{\partial \varphi} F_{0 x} F^{0 x} +  \frac{1}{8\pi} \frac{\partial (\lambda + \lambda_{B}/2)}{\partial \varphi} F_{y z} F^{y z}=0,
\label{phineq1}\\
&& \dot{\rho} + ( H + 2 F) (\rho + p)=0,
\label{rhoneq1}
\end{eqnarray}
where $A_{0}$, $A_{\parallel}$ and $A_{\perp}$ are defined as:
\begin{eqnarray}
A_{0} &=& -  \frac{\lambda + \lambda_{E}}{8 \pi} F_{0 x} \, F^{0 x} + \frac{\lambda - \lambda_{B}}{8 \pi} F_{y z} F^{y z} = \frac{(\lambda + \lambda_{E})}{(\lambda + \lambda_{E}/2)^2} \rho_{E} + ( \lambda - \lambda_{B}) \rho_{B},
\label{A0}\\ 
 A_{\parallel} &=& - \frac{\lambda}{8 \pi} F_{0x} \, F^{0 x} + \frac{\lambda}{8 \pi} F_{y z} \, F^{y z} =  \frac{\lambda }{(\lambda + \lambda_{E}/2)^2} \rho_{E} + \lambda  \rho_{B},
 \label{Apar}\\
 A_{\perp} &=&   \frac{\lambda + \lambda_{E}}{8 \pi} F_{0 x} \, F^{0 x} - \frac{\lambda - \lambda_{B}}{8 \pi} F_{y z} F^{y z} = -\frac{(\lambda + \lambda_{E})}{(\lambda + \lambda_{E}/2)^2} \rho_{E} - ( \lambda - \lambda_{B}) \rho_{B}.
 \label{Aperp}
\end{eqnarray}
Introducing the shifted variables $\Lambda_{E} = \lambda + \lambda_{E}/2$ and 
$\Lambda_{B} = \lambda + \lambda_{B}/2$, Eq. (\ref{phineq1}) can be expressed as: 
\begin{equation}
\ddot{\varphi} + ( H + 2 F) \dot{\varphi} + \frac{\partial V}{\partial \varphi} - \frac{\partial\Lambda_{E}}{\partial \varphi} \frac{\rho_{E}}{\Lambda_{E}^2}  +  \frac{\partial\Lambda_{B}}{\partial \varphi} \rho_{B}=0.
\label{KG1}
\end{equation}
Furthermore the analogs of Eqs. (\ref{ham1}) 
and (\ref{sh2}) become
\begin{eqnarray}
&&2 (\dot{H} + 2 \dot{F}) + ( H^2 + 2 F^2) + ( H+ 2 F)^2 =   \frac{1}{\overline{M}_{\mathrm{P}}^2} \biggl( 4 V - \dot{\varphi}^2 + \rho - 3 p - A_{0} - A_{\parallel} - 2 A_{\perp} \biggr). 
\nonumber\\
&& \dot{H} - \dot{F}  + (H - F) (H+ 2 F) = - \frac{1}{\overline{M}_{\mathrm{P}}^2} (A_{\parallel} - A_{\perp}).
\label{sh2a}
\end{eqnarray}
Finally, eliminating $\dot{H}$ and $\dot{F}$ from Eq. (\ref{00neq1}) with the help of Eqs. (\ref{xxneq1}) and (\ref{yyneq1}) we obtain 
the explicit form of the Hamiltonian constraint:
\begin{equation}
(H^2 + 2 F^2) - (H+ 2F)^2 = - \frac{2}{\overline{M}_{\mathrm{P}}^2 } \biggl[V + \frac{\dot{\varphi}^2}{2} + \rho + \frac{A_{0} - A_{\parallel} - A_{\perp}}{2} \biggr].
\label{ham2}
\end{equation}
\subsection{Constant curvature solutions}
Let us look for solutions of the system characterized by constant space-time curvature (i.e. $\dot{H} =0$,  $\dot{F} =0$), linear inflaton (i.e. $\ddot{\varphi}=0$) and in the absence of fluid sources  (i.e. $\rho=0$ and $p =0$).  Equations (\ref{xxneq1}) and (\ref{yyneq1})  reduce then to the following pair of algebraic conditions:
\begin{equation}
(H + 2 F)^2 =  \frac{1}{\overline{M}_{\mathrm{P}}^2} (3 V - A_{\parallel} - 2 A_{\perp}), \qquad (H - F) ( H + 2 F) = -  \frac{1}{\overline{M}_{\mathrm{P}}^2} ( A_{\parallel} - A_{\perp}).
\label{CC1}
\end{equation}
The solution of Eq. (\ref{CC1})  is:
\begin{equation}
H = \frac{1}{\overline{M}_{\mathrm{P}}}\frac{(V - A_{\parallel})}{\sqrt{3 V - A_{\parallel} - 2 A_{\perp}}}, \qquad F =\frac{1}{\overline{M}_{\mathrm{P}}} \frac{(V - A_{\perp})}{\sqrt{3 V - A_{\parallel} - 2 A_{\perp}}}.
\label{CC2}
\end{equation}
Taking the difference between Eqs. (\ref{00neq1}) and (\ref{xxneq1}) we can deduce $\dot{\varphi}^2$
\begin{equation}
 \dot{\varphi}^2 =  - 2 \overline{M}_{\mathrm{P}}^2 \,F \,( F - H)  + A_{\parallel} -  A_{0}.
\label{CC3}
\end{equation}
Using Eq. (\ref{CC2}) into Eq. (\ref{CC3}) we obtain 
\begin{equation}
 \dot{\varphi}^2 =  - A_{0} + \frac{(V - A_{\parallel}) A_{\parallel}}{3 V - A_{\parallel} - 2 A_{\perp}}  + \frac{2 (V - A_{\perp}) A_{\perp}}{3 V - A_{\parallel} - 2 A_{\perp}}.
\label{CC4}
\end{equation} 
In the case of magnetic initial conditions $\Lambda_{B}$ and $\lambda_{B}$ can be parametrized, for instance, as
\begin{equation}
\Lambda_{B} = \Lambda_{B\,0} \exp{[ \delta \varphi/\overline{M}_{\mathrm{P}}]}, \qquad \lambda_{B} = \lambda_{B\,0} \exp{[ \delta \varphi/\overline{M}_{\mathrm{P}}]}.
\label{CC5}
\end{equation}
Furthermore, the potential and $\dot{\varphi}$ can then be written as
\begin{equation}
V = A_{\perp} - \frac{\delta^2}{4} \overline{\rho}_{B} \Lambda_{B0},\qquad \dot{\varphi} = \frac{4 F \overline{M}_{\mathrm{P}}}{ \delta}.
\label{CC6}
\end{equation}
As a consequence of Eqs. (\ref{CC5}) and (\ref{CC6}) we have 
\begin{eqnarray}
A_{0}&=& (\lambda_{0} - \lambda_{B 0}) \overline{\rho}_{B} = (\Lambda_{B0} - 3 \lambda_{B0}/2)  \overline{\rho}_{B},
\nonumber\\
A_{\parallel} &=&  (\Lambda_{B0} -  \lambda_{B0}/2)  \overline{\rho}_{B},
\nonumber\\
A_{\perp}&=&- (\lambda_{0} - \lambda_{B 0})  \overline{\rho}_{B}= -(\Lambda_{B0} - 3 \lambda_{B0}/2)  \overline{\rho}_{B},
\label{CC7}
\end{eqnarray}
where it is practical to introduce the magnetic energy density $\overline{\rho}_{B} = B^2/8\pi$.

The combination of Eq. (\ref{CC6}) with Eq. (\ref{CC4}) fixes the value of $\delta$: 
\begin{equation}
\delta^2 = \frac{8}{7} \biggl(\frac{\lambda_{B0}}{\Lambda_{B0}} -1\biggr), \qquad \frac{\lambda_{B0}}{\Lambda_{B0}} > 1.
\label{CC8}
\end{equation}
But since $\delta^2> 0$ and $\lambda_{B0} > \Lambda_{B0}$ also the potential is positive definite, as expected:
\begin{equation}
V = \frac{\overline{\rho}_{B}}{14} ( 17 \lambda_{B0} - 10 \Lambda_{B0} ) >0.
\label{CC9}
\end{equation}
With the value of $\delta^2$ determined by Eq. (\ref{CC8}) the explicit numerical values of $H$ and $F$ is 
\begin{equation}
H= \frac{4}{\sqrt{14}} \frac{\sqrt{\overline{\rho}_{B}}}{\overline{M}_{\mathrm{P}}} \sqrt{ \lambda_{B0} - \Lambda_{B0}}, \qquad 
F = -  \frac{1}{\sqrt{14}} \frac{\sqrt{\overline{\rho}_{B}}}{\overline{M}_{\mathrm{P}}} \sqrt{ \lambda_{B0} - \Lambda_{B0}}.
\label{CC10}
\end{equation}
Note that $H>0$ but $F<0$; more importantly $(H + 2 F) > 0$ so that the background is, in average, inflating as 
expected. 

\subsection{Power law solutions}

We can investigate the power law solutions by inserting the analog of Eq. (\ref{ans1}) into Eqs. (\ref{00neq1})--(\ref{yyneq1}). Summing up term by term Eqs. (\ref{00neq1}) and (\ref{yyneq1}) the resulting equation only contains the scalar potential:
\begin{equation}
(\alpha + \beta) ( \alpha + 2 \beta -1) = \frac{2 \, t^2\, V}{\overline{M}_{\mathrm{P}}^2} - t^2 \frac{(A_{\perp} + A_{\parallel})}{\overline{M}_{\mathrm{P}}^2},
\label{con1neqa}
\end{equation}
Subtracting Eq. (\ref{xxneq1}) from Eq. (\ref{00neq1}), $\dot{\varphi}^2$ is then determined from the following relation:
\begin{equation}
\beta( \beta - \alpha -1) = - \frac{\dot{\varphi}^2 t^2}{2 \overline{M}_{\mathrm{P}}^2} - \frac{t^2 (A_{0} - A_{\parallel})}{2 \overline{M}_{\mathrm{P}}^2}. 
\label{con2neqa}
\end{equation}
In the limit $\lambda_{B} \to 0$ and $\lambda_{E} \to 0$ the last terms on the right hand side of Eqs. 
(\ref{con1neqa}) and (\ref{con2neqa}) vanish. If only $\lambda_{B}$ (or $\lambda_{E}$) vanishes, the supplementary 
contributions in Eqs. (\ref{con1neqa}) and (\ref{con2neqa}) do not disappear and the situation is still different. 
Subtracting Eq. (\ref{yyneq1}) from Eq. (\ref{xxneq1}) we obtain the further condition 
\begin{equation}
(\alpha - \beta) ( \alpha + 2 \beta -1) = \,\frac{2 t^2 (A_{\perp} - A_{\parallel})}{\overline{M}_{\mathrm{P}}^2}.
\label{con3neqa}
\end{equation}

As in the case of constant curvature solutions let us posit that $V(\varphi)$ has an exponential form (i.e. $V = V_{0} \exp{[\gamma \varphi/\overline{M}_{\mathrm{P}}]}$) so that, for consistency,  also  $\Lambda_{B}(\varphi)$ must be an exponential of $\varphi$:
\begin{equation}
 \Lambda_{B}(\varphi) = \Lambda_{B0} \exp{[\delta_{B} \, \varphi/\overline{M}_{\mathrm{P}}]}, \qquad \Lambda_{E}(\varphi) = \Lambda_{E0} \exp{[\delta_{E} \, \varphi/\overline{M}_{\mathrm{P}}]}.
\label{con4neq}
\end{equation}
We shall also assume, for sake of simplicity, that $\lambda_{B}$ and $\lambda_{E}$ have the same dependence of $\varphi$ but with a different overall normalization, i.e. 
$\lambda_{B}/\Lambda_{B} = \lambda_{B0}/\Lambda_{B0}$ and $\lambda_{E}/\Lambda_{E} = \lambda_{E0}/\Lambda_{E0}$.
In the case of magnetic initial conditions, Eqs. (\ref{con1neqa}), (\ref{con2neqa}) and (\ref{con3neqa}) imply the following set of algebraic conditions on the parameters of the solution:
\begin{eqnarray}
&& \Omega_{B} \lambda_{B0} - \varphi_{1}^2 = 2 \beta( \beta - \alpha - 1),
\label{alB1}\\
&& (\alpha - \beta) ( \alpha + 2 \beta -1) = 2 (\lambda_{B0} - \Lambda_{B0}) \Omega_{B},
\label{alB2}\\
&& (\alpha + \beta) ( \alpha + 2 \beta -1) = \frac{2 V_{0}}{\overline{M}_{\mathrm{P}}^2} -  \lambda_{B0}\Omega_{B}.
\label{alB3}
\end{eqnarray}
Equations (\ref{alB1})--(\ref{alB3})  must be complemented by the following pair of relations involving $\delta_{B}$ and $\gamma$:
\begin{equation}
2 - 4 \beta + \delta_{B} \varphi_{1} =0, \qquad \varphi_{1} = - \frac{2}{\gamma}.
\label{alB4}
\end{equation}
The algebraic relations obtained stemming from Eqs. (\ref{alB1})--(\ref{alB4}) must be consistent with Eq. (\ref{KG1}) which is 
solved only if $(\alpha + 6 \beta - 4) =0$. If the geometry globally expands we must have that $(\alpha + 2 \beta)>0$. But this implies that $\beta < 1$ since $\alpha + 2 \beta = 4 ( 1 -\beta)$. 

In the case of electric initial conditions Eqs. (\ref{alB1})--(\ref{alB4}) are modified as follows:
\begin{eqnarray}
&& \Omega_{E} \frac{\lambda_{E0}}{\Lambda_{E0}^2} - \varphi_{1}^2 = 2 \beta( \beta - \alpha - 1),
\label{alE1}\\
&& (\alpha - \beta) ( \alpha + 2 \beta -1) = - 2 \frac{\Omega_{E}}{\Lambda_{E0}},
\label{alE2}\\
&& (\alpha + \beta) ( \alpha + 2 \beta -1) = \frac{2 V_{0}}{\overline{M}_{\mathrm{P}}^2} + \frac{\lambda_{E0}}{\Lambda_{E0}^2}\Omega_{B}, 
\label{alE3}\\
&& 2 - 4 \beta - \delta_{E} \varphi_{1} =0,
\label{alE4}
\end{eqnarray}
while it is still true that $\varphi_{1} = - 2/\gamma$. The consistency 
of electric initial conditions with Eq. (\ref{KG1}) demands that $(\alpha + 2 \beta) = 2$ which is always positive. Furthermore, since $\Omega_{E} > 0$, Eq. (\ref{alE2}) implies $\beta> 3/2$.

\subsection{Interpolating solutions}

In the case where the susceptibilities do not coincide it is possible to find solutions interpolating 
between the power-law and the constant curvature regimes. These solutions may 
describe the protoinflationary evolution, i.e. the transition between an expanding epoch and the accelerated evolution.

Let us therefore focus on the case of magnetic initial conditions and construct the solutions 
by using the results obtained so far. As before we shall assume 
that $\lambda_{B}$ and $\lambda$ depend exponentially  on $\varphi$ but with different 
parameters i.e. 
\begin{equation}
\lambda_{B}(\varphi) = \lambda_{B 0} \exp{[\delta_{1}  \varphi/\overline{M}_{\mathrm{P}}]}, \qquad 
\lambda(\varphi) = \lambda_{ 0} \exp{[\delta_{2}  \varphi/\overline{M}_{\mathrm{P}}]}.
\label{int2a}
\end{equation}
The equation determining $\dot{\varphi}$ can be written is, in this case,
\begin{equation}
\dot{F} + F ( F - H) = - \frac{1}{2 \overline{M}_{\mathrm{P}}^2} [ \dot{\varphi}^2 + ( p + \rho) + ( A_{0} - A_{\parallel})].
\label{FF}
\end{equation}
From Eq. (\ref{int1}) we can solve Eq. (\ref{rhoneq1}) giving the evolution of $\rho$, i.e.
\begin{equation}
\rho(t) = \rho_{*} a^{- (w + 1)} \, b^{- 2 (w+1)}, \qquad p = w \,\rho,
\label{int2} 
\end{equation}
where $w$ denotes the barotropic index of the fluid. Inserting Eq. (\ref{int2})  into Eq. (\ref{FF}) and recalling the explicit expressions for $A_{0}$ and $A_{\parallel}$ we have;
\begin{equation}
\frac{\partial}{\partial t} \biggl(\frac{b}{a} F\biggr) = - \frac{b}{ 2 \overline{M}_{\mathrm{P}}^2 a} \biggl[  \dot{\varphi}^2
+ \frac{\rho_{*} (w + 1)}{( a\, b^2)^{w +1}}  - \lambda_{B} \frac{ B^2}{8 \pi b^4} \biggr].
\label{FF2}
\end{equation}
Equation (\ref{FF2}) admits a particular solution with the correct asymptotic behaviour, namely, 
\begin{equation}
a(t) = \bigl[\sinh{(H_{*} t)}\bigr]^{\alpha}, \qquad b(t) = \bigl[\sinh{(H_{*} t)}\bigr]^{\beta},
\label{int1}
\end{equation}
interpolating between a power-law phase (for $H_{*}\, t < 1$) and a constant curvature regime (for $H_{*} \,t > 1$) that can be inflating 
provided $\alpha + 2 \beta >0$. The scalar field $\varphi(t)$ is then determined by imposing the restriction
$0 < \beta < \alpha$:
\begin{equation}
\varphi(t) = \overline{M}_{\mathrm{P}} \, \sqrt{ 2 \beta (\alpha - \beta)} \, \ln{[ \sin{(H_{*} \, t)}]}.
\label{int3}
\end{equation}
Equations (\ref{FF2}) and (\ref{int1})--(\ref{int3}) demand three specific relations 
among the parameters, namely:
\begin{equation}
\delta_{1} = \sqrt{\frac{2}{\beta(\beta - \alpha)}} (2 \beta -1),\qquad 
\beta = \frac{w +1}{2} \Omega_{*} - \frac{\lambda_{B0}}{2} \Omega_{B}, \qquad (\alpha + 2 \beta) (w+1) =2,
\label{int5}
\end{equation}
where $\rho_{*} = H_{*}^2 \overline{M}_{\mathrm{P}}^2\, \Omega_{*}$ and 
$\overline{\rho}_{B} =\Omega_{B}  H_{*}^2 \overline{M}_{\mathrm{P}}^2$ (recall that $\overline{\rho}_{B} = B^2/8\pi$ and $\rho_{*}$ are 
constants so that also $\Omega_{*}$ and $\Omega_{B}$ are constant parameters). The obtained solution leads to a consistent determination of the scalar potential $V(\varphi)$. In particular, by combining Eqs. (\ref{xxneq1}) and (\ref{yyneq1}), the following relation can be obtained:
\begin{equation}
(\dot{H} + \dot{F}) + ( H + F) ( H + 2 F) = - \frac{1}{\overline{M}_{\mathrm{P}}^2} [ (p - \rho) - 2 V +  A_{\parallel} + A_{\perp}].
\label{FF3}
\end{equation}
which can also be written as: 
\begin{equation}
\frac{\partial }{\partial t} [ a\, b^2 \, ( H + F) ] = - \frac{a\, b^2}{\overline{M}_{\mathrm{P}}^2}[ (p - \rho) - 2 V + A_{\parallel} + A_{\perp}].
\label{FF4}
\end{equation}
Inserting Eq. (\ref{int1}) into Eq. (\ref{FF4}) and recalling Eq. (\ref{int3}), the potential $V(\varphi)$ becomes:
\begin{eqnarray}
V(\varphi) &=& \overline{V}_{1} + \overline{V}_{2} \exp{[- \delta_{3}\varphi/\overline{M}_{\mathrm{P}}]},\qquad \delta_{3} = \sqrt{\frac{2}{\beta (\alpha -\beta)}},
\nonumber\\
\overline{V}_{1} &=& \frac{\overline{M}_{\mathrm{P}}^2 \, H_{\star}^2}{2} (\alpha + \beta) ( \alpha + 2 \beta),
\nonumber\\
\overline{V}_{2} &=& \frac{\overline{M}_{\mathrm{P}}^2 \, H_{\star}^2}{2}[(\alpha+ \beta) (\alpha + 2 \beta-1)  + (w -1) \Omega_{*} + \lambda_{B0} \Omega_{B} ].
\label{int6}
\end{eqnarray}
The shear equation (obtained by subtracting Eq. (\ref{yyneq1}) from Eq. (\ref{xxneq1}))
\begin{equation}
\frac{\partial}{\partial t}  [ a\, b^2 \, ( H - F) ] = -  \frac{a \, b^2 }{ \overline{M}_{\mathrm{P}}^2} ( 2 \lambda - \lambda_{B}) \rho_{B},
\label{FF5}
\end{equation}
is  automatically satisfied provided the following conditions hold:
\begin{equation}
2 \lambda_{0} \Omega_{B} = - (\alpha -\beta)(\alpha + 2 \beta), \qquad \lambda_{B0} \, \Omega_{B} = (\alpha - \beta) ( \alpha + 2\beta -1),
\label{FF6}
\end{equation}
from which it is clear that $\lambda_{0}$ and $\lambda_{B0}$ must have opposite sign. 
Equation (\ref{FF5}) also implies, for consistency, $\delta_{2} = 4 \beta/\sqrt{2 \beta( \alpha - \beta)}$. 
Inserting the obtained solution into Eq. (\ref{phineq1}) and using the relations among the various parameters we have 
that the equation for $\varphi$ (i.e. Eq. (\ref{KG1})) is satisfied provided
\begin{equation}
(\alpha + 2 \beta -1)\, (\alpha - \beta)\, ( \alpha + 6 \beta - 4) =0.
\label{int7}
\end{equation}
This condition (\ref{int7}) is trivially satisfied when $\alpha = \beta$ (i.e. the isotropic case). In the isotropic case 
this kind of protoinflationary solution has been recently discussed in a related context \cite{proto2,proto1}.
The condition (\ref{int7}) can also be satisfied if $\alpha + 2 \beta = 1$ or if $\alpha + 2 \beta = 4 ( 1 -\beta)$. 
Since we must impose $(\alpha + 2 \beta) >0$ (to have inflation) and $\alpha > \beta$ (for algebraic consistency),
$0< \beta<4/7$.These two conditions correspond, in the present example, to different protoinflationary evolutions. 
The same kind of solutions can be investigated in the electric case when $\lambda_{B} =0$ but $\lambda_{E} \neq 0$.
In the electric case, however, the set of algebraic conditions cannot be analytically satisfied and will not be discussed 
any further.

\renewcommand{\theequation}{5.\arabic{equation}}
\setcounter{equation}{0}
\section{Protoinflationary dynamics and stability}
\label{sec5}
\subsection{General considerations}
Quasi-de Sitter expansion can last more than $63$ efolds\footnote{For the fiducial set of parameters of the concordance scenario (see e.g. \cite{wmap1,wmap2,wmap3}), the maximal number of inflationary efolds accessible to large-scale observations is ${\mathcal O}(63)$ \cite{eff}.} but it cannot continue 
indefinitely in the past because of the lack of geodesic completeness of the conventional inflationary backgrounds. The standard phase of accelerated 
expansion is customarily complemented by a preinflationary stage where the total energy-momentum tensor is dominated by a globally neutral plasma \cite{nonvac} and where the scale factor expands but in a decelerated manner. The protoinflationary epoch of expansion coincides approximately with the end of the preinflationary time
when the background geometry starts accelerating. 

If the gauge hairs determine the initial conditions of the shear parameter $r$, the features of the preinflationary stage  are essential for the survival of a given solution. Electric and magnetic hairs have different interactions with the ambient plasma: while electric fields are likely to be screened and dissipated, magnetic fields can be present and stable in ideal conductors. The simplest physical realization of the 
protoinflationary plasma is represented by a globally neutral system containing both charged species and neutral species \cite{proto2,proto1}. Moreover the evolution of the geometry is  very similar to the scenario suggested by the exact solution of Eq. (\ref{int1}) where for $t \ll H_{*}^{-1}$ the background decelerates while for $t\gg H_{*}^{-1}$ the background inflates. In this example, which has an isotropic counterpart, $H_{*}^{-1}$ marks the time-scale of the protoinflationary dynamics.

In an expanding background the screening properties of the plasma are always controlled by the ratio ${\mathcal N}_{r}/{\mathcal N}_{0}$ \cite{proto2,proto1} where ${\mathcal N}_{r}$ denotes the concentration of neutral species (e. g. photons) while ${\mathcal N}_{0}$ is  the common concentration of positive and negative charge carriers. In the realistic situation (i.e. ${\mathcal N}_{r} \gg {\mathcal N}_{0}$) the temperature of the charged species approximately coincides with the one of the neutrals and the electric fields are screened (depending on the smallness of the plasma parameter) exactly as it happens in laboratory plasmas (see e.g. \cite{KR}).  The contribution of the charged species to the transport coefficients can then be computed assuming that the collisions between the particles of the same charge are negligible as it happens for Lorentzian plasmas \cite{KT}.  Even if the specific value of the transport 
coefficients depends on the microscopic model of charge carriers, the conductivity of the protoinflationary plasma scales approximately as $T$, i.e. the 
temperature $T$ of the dominant neutral species. As soon as the temperature drops below the value of the mass of the lightest charge carrier 
the conductivity will scale as $T^{2/3}$. Thus the a realistic model of protoinflationary conductivity implies that $\sigma$ interpolates between $T$ and $T^{3/2}$ \cite{proto2}. In this case the various energy densities 
of the plasma will evolve as 
\begin{eqnarray}
&& \dot{\rho}_{B} + 4 n \biggl( 1 - \frac{r}{3} \biggr) \rho_{B}=0,\qquad \dot{\rho} + 3 n ( \rho + p) =0,
\label{rB}\\
&& \dot{\rho}_{E} + 4 n \biggl( 1 - \frac{r}{3} \biggr) \rho_{E} + \frac{8\pi \sigma}{\Lambda_{E}} \rho_{E} = 0,
\label{rE}\\
&& \dot{\rho}_{\varphi} + 3 n ( \rho_{\varphi} + p_{\varphi}) + \rho_{B} \dot{\Lambda}_{B}  - \frac{\rho_{E}}{\Lambda_{E}^2} \dot{\Lambda}_{E} =0.
\label{rphex}
\end{eqnarray}
Equations (\ref{rB})--(\ref{rphex}) can be solved numerically together with the equations for $n$ and $r$, i.e. 
\begin{eqnarray}
&& \dot{n} + 2 n^2 \biggl( 1 + \frac{r^2}{18}\biggr) = - \frac{{\mathcal D}_{n}}{6 \overline{M}_{\mathrm{P}}^2},
\label{ngen}\\
&& \dot{r} + n \, r \biggl(1 - \frac{r^2}{9} \biggr) = - \frac{( A_{\parallel} - A_{\perp})}{n\,\overline{M}_{\mathrm{P}}^2} + \frac{r}{ 6\, n\, \overline{M}_{\mathrm{P}}^2} {\mathcal D}_{n},
\label{rgen}
\end{eqnarray}
where ${\mathcal D}_{n}$ is defined as:
\begin{eqnarray}
{\mathcal D}_{n} &=& (3 p_{\varphi} - \rho_{\varphi}) + (3 p - \rho) + A_{0} + A_{\parallel} + 2 A_{\perp}.
\label{DN}
\end{eqnarray}

Instead of going through a series of detailed numerical examples, we prefer to 
solve approximately the previous system of equations in the regime where the inflaton is not yet dominant and the geometry is dominated by radiation (i.e. $(3 p - \rho)\simeq 0$). In this scheme the dimensionless shear parameter is much smaller than one (i.e. $r \ll 1$) and  Eqs. (\ref{ngen}) and (\ref{rgen}) 
can be perturbatively solved by using $r$ as the expansion parameter:
\begin{equation}
\dot{n} + 2 n^2 \simeq 0, \qquad \dot{r} + n r \simeq -  \frac{1}{n\,\overline{M}_{\mathrm{P}}^2}( A_{\parallel} - A_{\perp}).
\label{exx1}
\end{equation}
Within the same approximation the solution of Eqs. (\ref{rB}) and (\ref{rE}) can be written as:
\begin{eqnarray}
\rho_{E}(t) &=& \rho_{E}(t_{i}) \biggl(\frac{n(t)}{n(t_{i})}\biggr)^2 e^{- {\mathcal J}(t_{i}, t)}, \qquad  {\mathcal J}(t_{i}, t) = \int_{t_{i}}^{t} \frac{8\pi \sigma(t')}{\Lambda_{E}(t')} \, d t' = {\mathcal O}\biggl(\frac{\sigma_{*}}{n_{*} \Lambda_{E*}}\biggr),
\label{exx2a}\\
\rho_{B}(t) &=& \rho_{B}(t_{i}) \biggl(\frac{n(t)}{n(t_{i})}\biggr)^2,
\label{exx2b}
\end{eqnarray}
where $\sigma(t_{*}) \simeq \sigma_{*}$ and $t_{i}$ denotes the initial integration time $t_{i} < t_{*}$. Since $\sigma_{*} \simeq {\mathcal O}(T_{*})$, in spite of the value of  $\Lambda_{E*}$  we have to admit that 
$\sigma_{*}/n_{*} \gg 1$ so that it is difficult to imagine situations where the electric fields are not exponentially 
suppressed by the finite value of the protoinflationary  conductivity. In practice only the magnetic fields will survive 
the protoinflationary phase. Thus the asymptotic value of the shear parameter which could be eventually used to construct
anisotropic inflationary solutions can be obtained by solving Eq. (\ref{exx1}) with the result that:
\begin{equation}
r(t) \simeq - 3 \alpha_{B} \omega_{B}(t_{i}) \biggl[ 1 - \sqrt{\frac{t_{i}}{t}}\biggr], \qquad t \leq t_{*}
\label{exx3}
\end{equation}
where $\omega_{B}(t) = \rho_{B}/( 3 n^2 \overline{M}_{P}^2)$; $\alpha_{B} \simeq (2 \lambda - \lambda_{B})$ and the susceptibilities have been assumed to vary slowly during the protoinflationary phase. 

\subsection{Autonomous systems} 
It would be tempting to conclude that the existence of magnetic hairs in a given background inflating anisotropically is just a sufficient requirement for the stability of the solution. This is not the case since two different solutions carrying magnetic hairs lead to anisotropic fixed points with opposite 
stability properties. This means that the existence of magnetic hairs is not sufficient for the dynamical stability 
of the corresponding solution but it is nonetheless necessary.

Recalling that  $\gamma = \mathrm{det}(\gamma_{ij})$, Eqs. (\ref{ngen}) and (\ref{rgen}) can be rephrased in terms of the variable $x = \ln{(\gamma^{1/6})}$
with the result that\footnote{ It follows from the definition of $x$ that $n = \dot{x}$.
The variable $x$ employed hereunder should not be confused with the homonymous  spatial coordinate that never appears explicitly in the remaining part of this section. In the case of the background of Eq. (\ref{FF0}) we have $x = \ln{(a b^2)^{1/3}}$, giving, 
in the isotropic limit $x\to \ln{a}$.}
\begin{eqnarray}
&& \frac{ d n}{d x} = - 2 n\biggl( 1 + \frac{r^2}{18} \biggr) - \frac{{\mathcal D}_{n}}{6 \, n\,\overline{M}_{\mathrm{P}}^2},
\label{nx}\\
&& \frac{d r}{dx} =  - r \biggl( 1 - \frac{r^2}{9}\biggr) - \frac{(A_{\parallel} - A_{\perp})}{n^2 \, \overline{M}_{\mathrm{P}}^2} + \frac{r }{ 6 \, n^2 \, \overline{M}_{\mathrm{P}}^2} {\mathcal D}_{n}.
\label{rx}
\end{eqnarray}
The Hamiltonian constraint of Eq. (\ref{ham2}) can be written in terms of the Zeldovich 
variables $n$ and $r$, i.e. 
\begin{equation}
 6 \overline{M}_{\mathrm{P}}^2 n^2 \biggl( 1 - \frac{r^2}{9} \biggr) = \dot{\varphi}^2 + 2 V + 2 \rho + A_{0} - A_{\parallel} - 2 A_{\perp}.
 \label{cons1}
 \end{equation}
After introducing the following dimensionless quantities:
\begin{equation}
\omega_{V} = \frac{V}{6 n^2 \overline{M}_{\mathrm{P}}^2}, \qquad \omega_{\rho} = \frac{\rho}{6 n^2 \overline{M}_{\mathrm{P}}^2}, \qquad 
p = \frac{1}{\overline{M}_{\mathrm{P}}} \biggl(\frac{d \varphi}{d x}\biggr),
\label{cons3}
\end{equation}
Eq. (\ref{cons1}) can be expressed as: 
\begin{equation}
\omega_{V} =  1 - \frac{r^2}{9} - \frac{p^2}{6} - \omega_{\rho} - \frac{ A_{0} - A_{\parallel} - 2 A_{\perp}}{ 6 \overline{M}_{\mathrm{P}}^2 n^2 }.
\label{cons2}
\end{equation}
Inserting Eq. (\ref{cons2}) into Eq. Eq. (\ref{nx})
and elminating $\omega_{V}$ from ${\mathcal D}_{n}$ the following equation can be obtained
\begin{equation}
\frac{d s}{dx} = - \biggl( \frac{r^2}{3} + \frac{p^2}{2}\biggr) - \frac{3}{2}(w + 1) \omega_{\rho} - \frac{3 A_{0} - ( A_{\parallel} + 2 A_{\perp})}{6 n^2 \overline{M}_{\mathrm{P}}^2},
\label{nx2}
\end{equation}
where $s = \ln{n}$. 
Equation (\ref{cons2}) can be used to eliminate $\omega_{V}$ from all 
the relevant equations. Thus, denoting with the prime a derivation with respect to $x$ and making explicit the dependence on the gauge fields, Eqs. (\ref{nx}) and (\ref{rx}) become, in the new parametrization:
\begin{eqnarray}
s' &=& -  \biggl( \frac{r^2}{3} + \frac{p^2}{2}\biggr) - \frac{3}{2} (w + 1) \omega_{\rho} - \frac{4 \lambda - 5 \lambda_{B}}{2} \omega_{B} - 
\frac{4 \lambda + 5 \lambda_{E}}{2 \Lambda_{E}^2} \omega_{E},
\label{s1}\\
r' &=& - 3 r \biggl( 1 - \frac{r^2}{9} - \frac{p^2}{6} \biggr)+ \frac{3}{2}\, r\,  (w + 1) \omega_{\rho} 
+ \omega_{E} \biggl( r\, \frac{4 \lambda+ 5 \lambda_{E}}{2 \Lambda_{E}^2} 
- \frac{6}{\Lambda_{E}} \biggr) 
\nonumber\\
&+& \omega_{B} \biggl[ r\, \frac{4 \lambda- 5 \lambda_{B}}{2} 
- 3 ( 2 \lambda - \lambda_{B}) \biggr],
\label{r1a}
\end{eqnarray}
where $\omega_{E}$ and $\omega_{B}$ are defined in full analogy with $\omega_{V}$ and $\omega_{\rho}$:
\begin{equation} 
\omega_{E} = \frac{ \rho_{E} }{ 3 \, n^2 \, \overline{M}_{\mathrm{P}}^2}, \qquad \omega_{B} = \frac{ \rho_{B} }{ 3 \, n^2 \, \overline{M}_{\mathrm{P}}^2}.
\label{r1b}
\end{equation}
Equation (\ref{s1}) is just a definition of the slow-roll parameter expressed in the language of the Zeldovich variables, as the following chain of equalities clearly shows:
\begin{equation}
\frac{d s}{d x} = \frac{n'}{n} =  \frac{\dot{n}}{n^2} = - \epsilon.
\label{SR}
\end{equation}

The equations for $\omega_{\rho}$,  $\omega_{B}$ and $\omega_{E}$ can be easily obtained and it is:
\begin{eqnarray}
\omega_{\rho}' &=& - 3 (w + 1) \omega_{\rho}  + {\mathcal G}(r,\,p,\,\omega_{\rho},\,\omega_{B},\, \omega_{E})\omega_{\rho},
\label{omr1}\\
\omega_{B}' &=&  -  4\biggl( 1 - \frac{r}{3}\biggr) \omega_{B} +  {\mathcal G}(r,\,p,\,\omega_{\rho},\,\omega_{B},\, \omega_{E}) \omega_{B},
\label{omegaB}\\
\omega_{E}' &=& - \frac{8 \pi \sigma_{c}}{n \Lambda_{E}}\, \omega_{E}  -  4\biggl( 1 - \frac{r}{3}\biggr) \omega_{E} +  {\mathcal G}(r,\,p,\,\omega_{\rho},\,\omega_{B},\, \omega_{E})\omega_{E},
\label{omegaE}
\end{eqnarray}
where 
\begin{equation}
 {\mathcal G}(r,\,p,\,\omega_{\rho},\,\omega_{B},\, \omega_{E}) =   \biggl(\frac{2}{3} r^2 + p^2 \biggr) + 3 ( w + 1) \omega_{\rho}
+ (4 \lambda - 5 \lambda_{B}) \omega_{B} - \frac{(4 \lambda + 5 \lambda_{E})}{\Lambda_{E}^2} \omega_{E}.
\label{calG}
\end{equation}
Finally the equation for $p$ is given by\footnote{The constant $\gamma$ appearing in Eqs. (\ref{peq})--(\ref{peq2}) and coming from the potential of the 
inflaton must not be confused with the determinant  of the spatial part of the metric.}:
\begin{eqnarray}
p' &=& - 3 ( p + \gamma) \biggl( 1 - \frac{r^2}{9} - \frac{p^2}{6} \biggr) + \frac{3}{2} [ (w+1) p + 2 \gamma] \omega_{\rho}
\nonumber\\
&+& ( p + 3 \gamma) \biggl[ \frac{(\lambda + 2 \lambda_{E})}{2 \Lambda_{E}^2} + \frac{(\lambda - 2 \lambda_{B})}{2} \omega_{B} \biggr]
\nonumber\\
&+& \frac{3}{2} (p +\gamma) \biggl[ \frac{(\lambda + \lambda_{E})}{\Lambda_{E}^2} \omega_{E} + (\lambda - \lambda_{B}) \omega_{B} \biggr] - 3 \delta_{B} \Lambda_{B} \omega_{B} + \frac{3}{\Lambda_{E}} \delta_{E} \omega_{E}.
\label{peq}
\end{eqnarray}
Equation (\ref{peq}) has been written in the case of an exponential potential and for exponential couplings $\Lambda_{B}$ and $\Lambda_{E}$, i.e. 
\begin{equation}
\frac{\partial V}{\partial \varphi} = \frac{\gamma}{\overline{M}_{\mathrm{P}}} \, V, \qquad \frac{\partial \Lambda_{B}}{\partial \varphi} = \frac{\delta_{B}}{\overline{M}_{\mathrm{P}}} \, \Lambda_{B},\qquad \frac{\partial \Lambda_{E}}{\partial \varphi} = \frac{\delta_{E}}{\overline{M}_{\mathrm{P}}} \, \Lambda_{E}.
\label{peq2}
\end{equation}

The system of Eqs. (\ref{s1})--(\ref{peq}) includes also the fluid sources 
possibly present during the protoinflationary phase. These sources shall now be neglected but can be 
taken into account in a more refined treatment which is beyond the scope of this analysis.  In the remaining part of this section the stability properties of two different magnetic fixed points derived in sections \ref{sec3} and \ref{sec4} shall be compared. With this exercise we ought to show that  the existence of magnetic hairs is a necessary requirement for the stability of the solution but is is 
not sufficient.
\subsection{Stability of the magnetic solutions}
Defining the fluctuations of $r$, $p$ and $\overline{\omega}_{B}$ as\footnote{It is 
technically useful to rewrite Eq. (\ref{omegaB}) by introducing the rescaled variable $\overline{\omega}_{B} = \lambda \omega_{B}$ and to focus the attention on the case 
where $\lambda_{B}$ is proportional to $\lambda$.}:
\begin{equation}
\delta r = f_{r},\qquad \delta p= g_{p},\qquad \delta \overline{\omega}_{B} = h_{b},
\end{equation} 
the equations describing the fluctuations of the system around a given fixed point can be written as:
\begin{eqnarray}
f_{r}' &=& A_{r} f_{r} + B_{r} g_{p} + C_{r}h_{b},
\nonumber\\
g_{p}' &=& A_{p} f_{r} + B_{p} g_{p} + C_{p} h_{b},
\nonumber\\
h_{b}' &=& A_{b} f_{r} + B_{b}(\epsilon,\zeta) g_{p} + C_{b} h_{b},
\label{stab}
\end{eqnarray}
where the nine functions of Eq. (\ref{stab}) define the entries of the $3\times3$ stability matrix and depend on the parameters of the solution.
The fixed point of the magnetic solution of section \ref{sec3} can be expressed in the  $(\epsilon,\, \zeta)$ plane as\footnote{See Eq. (\ref{conn2}) and discussion therein.}:
\begin{eqnarray}
&& p\to p_{*}(\epsilon,\zeta) = - \sqrt{\frac{2 \, \epsilon}{3} } \, \sqrt{( 1 - \zeta) ( 3 + \epsilon\, \zeta)},
\nonumber\\
&&\gamma \to \gamma_{*}(\epsilon,\zeta) = \frac{\sqrt{6\, \epsilon}}{(1 - \zeta) ( 3 + \epsilon\, \zeta)},
\nonumber\\
&& \delta_{B} \to \delta_{*}(\epsilon,\zeta) = - \sqrt{\frac{2}{3 \epsilon}} \, \frac{6 - \epsilon( 3 - 2 \zeta) }{\sqrt{( 1 - \zeta) ( 3 + \epsilon \zeta)}}, 
\nonumber\\
&& \overline{\omega}_{B} \to \overline{\omega}_{B*}(\epsilon,\zeta) = \frac{\epsilon\, \zeta}{6} ( 3 - \epsilon),
\nonumber\\
&& r\to r_{*}(\epsilon,\zeta) = - \epsilon  \zeta.
\label{fixed}
\end{eqnarray}
The system of Eqs. (\ref{r1a}), (\ref{omegaB}) and (\ref{peq}) can then be perturbed around the fixed point given in Eq. (\ref{fixed}) and the 
nine functions appearing in Eq. (\ref{stab}) are:
\begin{eqnarray}
A_{r}(\epsilon,\zeta) &=& - 3 + \epsilon + \frac{2}{3} \epsilon^2 \zeta^2,
\nonumber\\
B_{r}(\epsilon,\zeta) &=&  \epsilon \,\sqrt{\epsilon} \,\zeta\, \sqrt{\frac{2}{3} } \sqrt{(1 - \zeta) ( 3 +\epsilon\zeta)},
\nonumber\\
C_{r}(\epsilon,\zeta) &=& - 2 \epsilon  \zeta -6,
\label{FR1}\\
A_{p}(\epsilon,\zeta) &=& \frac{2\sqrt{2}}{3 \sqrt{3}} \frac{( - 3 + \epsilon - \epsilon\,\zeta)}{\sqrt{(1 - \zeta)\, (3 + \epsilon \zeta)}}\, \epsilon \sqrt{\epsilon} \zeta^2 , 
\nonumber\\
 B_{p}(\epsilon,\zeta) &=& - 3 + \epsilon - 2 \epsilon \zeta + \frac{2}{3} \epsilon^2 \zeta ( 1 - \zeta),
\nonumber\\
 C_{p}(\epsilon,\zeta) &=& - \frac{2 \sqrt{2}}{\sqrt{3}\, \sqrt{\epsilon} ( 1 - \zeta)}\,  [\epsilon(1 - \zeta)- 3]\,\sqrt{(1 - \zeta)( 3 + \epsilon \zeta)} ,
\label{FR2}\\
A_{b}(\epsilon,\zeta) &=& - \frac{2\epsilon\, \zeta}{9} ( - 3 + \epsilon) ( 1 - \epsilon\zeta), 
\nonumber\\
B_{b}(\epsilon,\zeta) &=& \frac{\zeta\sqrt{\epsilon}}{3\sqrt{6}}
\frac{ (3 - \epsilon) \{ 6 + \epsilon[3 + 2\zeta( \epsilon(1 - \zeta)-2 )]\}}{\sqrt{(1 - \zeta) ( 3 + \epsilon\zeta)}}, 
\nonumber\\
C_{b}(\epsilon,\zeta) &=&  \frac{ 2 \epsilon \zeta}{3}(  3 - \epsilon).
\label{FR3}
\end{eqnarray}
The eigenvalues of the system can then be easily obtained and they are determined by the following equation:
\begin{equation}
( 3 - \epsilon + \mu)\, [ 3 \mu^2 ( 1 - \zeta) - 3 \mu ( - 3 + \epsilon) ( 1 - \zeta) + 2 \zeta (3 - \epsilon) (6 - \epsilon\, \zeta)] =0.
\label{FR4}
\end{equation}
Equation (\ref{FR4}) has three different roots:
\begin{eqnarray}
\mu_{\pm} &=& \frac{- 9 + 3 \epsilon + 9 \zeta - 3 \epsilon \zeta \pm \sqrt{3 ( \epsilon - 3) ( 1 -\zeta) [ 3 (\epsilon - 3) ( 1 - \zeta) + 8 \zeta ( 6 - \epsilon \zeta)]}}{6 ( 1 -\zeta)},
\nonumber\\
\mu_{0} &=& - 3 + \epsilon.
\label{FR5}
\end{eqnarray}
While it is immediately clear that, up to corrections ${\mathcal O}(\epsilon)$, $\mu_{0}$ is negative,  it is practical to expand the two roots $\mu_{\pm}$ it is in powers of $\zeta$ and $\epsilon$ which are, by definition, always small and positive; from Eq. (\ref{FR5}) the result 
of this expansion is:
\begin{eqnarray}
\mu_{+} &=& - 4 \zeta \biggl[ 1 + \frac{7}{3} \zeta +\frac{\zeta \epsilon}{18} + {\mathcal O}(\zeta^2) + {\mathcal O}(\zeta \epsilon^2)\biggr],
\nonumber\\
\mu_{-} &=&  - 3  + 4 \zeta + \epsilon + {\mathcal O}(\zeta^2) + {\mathcal O}(\zeta^2 \epsilon) + {\mathcal O}(\epsilon^2).
\label{FR6}
\end{eqnarray}
All the three roots are negative so that, in the language of the autonomous systems \cite{grim},  the obtained eigenvalues describe a stable node and the solution is overall stable.

A simplified description of the stability can be obtained by recalling that Eq. (\ref{SR}) can be expressed as $\epsilon = r^2/3 + p^2/2 + 2 \overline{\omega}_{B}$ (since $s'= - \epsilon$). 
The equation for $r$ and $p$ can be first written as: 
\begin{eqnarray}
&& r'= - ( 3 - \epsilon)r\, - 6 \overline{\omega}_{B}, 
\nonumber\\
&& p' = - (3 -\epsilon) (p + \gamma) + \gamma \overline{\omega}_{B}.
\label{SIMP}
\end{eqnarray}
By now perturbing around the fixed point, we can also safely assume $\delta \epsilon \simeq 0$ which implies, following the previous notations, that $2 f_{r} r/3 + p g_{p} + 2 h_{b} =0$.
Using the last relation, $h_{b}$  can be eliminated from the perturbed version of Eq. (\ref{SIMP}) 
and the system becomes:
\begin{eqnarray}
f_{r}' &=& - (3 - \epsilon) f_{r} + 2 r_{*} f_{r} + 3 p_{*} g_{p},
\nonumber\\
g_{p}' &=& - \gamma_{*} r_{*} f_{r}/3 - (3 - \epsilon + \gamma_{*} p_{*}/2) g_{p}.
\end{eqnarray}
The eigenvalue equation has two roots, i.e. 
$\mu_{\pm} = - 3 + 3 \epsilon/2 \pm \epsilon/2$ which are consistent with the existence of a stable 
node as already established. 

Let us now analyze another set of solutions with magnetic hairs, namely the constant curvature solutions derived in section \ref{sec4} (see Eqs. (\ref{CC8})--(\ref{CC10}) and discussion therein). The stability matrix depends now only on one parameter namely $\xi = \Lambda_{B0}/\lambda_{B0}$. The coefficients appearing in Eq. (\ref{stab}) only depend on $\xi$ and are defined as:
\begin{eqnarray}
A_{r}(\xi) &=& \frac{1704 + 96 \sqrt{14} - 1697 \xi}{32 ( 1 - \xi)},\qquad 
 B_{r}(\xi) = \frac{15 \sqrt{7}}{8\sqrt{2}} \sqrt{\frac{\xi}{1 - \xi}}, \qquad 
C_{r}(\xi) = 9,
\label{FFR1}\\
A_{p}(\xi) &=& - \frac{5 \sqrt{7}}{4 \sqrt{2}} \sqrt{\frac{\xi}{1 - \xi}},
\qquad 
B_{p}(\xi) = \frac{3( 336 + 64 \sqrt{14} - 329 \xi)}{64(1 - \xi)} ,
\nonumber\\
C_{p}(\xi) &=& -  \frac{ 6 \sqrt{2}}{\sqrt{7}} \sqrt{\frac{1 - \xi}{\xi}} - \frac{\sqrt{7}}{2 \sqrt{2}}\sqrt{ \frac{\xi}{1 - \xi}},
\label{FFR2}\\
A_{b}(\xi) &=& \frac{17 \sqrt{14}}{1 - \xi}, 
B_{b}(\xi) = \frac{3}{4 ( 1 - \xi)} \biggl(8 \sqrt{\frac{1 - \alpha}{\alpha}} - 7 
\sqrt{\frac{\xi}{1 - \xi}}\biggr), 
\nonumber\\
C_{b}(\xi) &=& \frac{6 \sqrt{14}}{1 - \xi}.
\label{FFR3}
\end{eqnarray}
In general terms the analysis of the roots of the corresponding eigenvalue equation is rather cumbersome. To avoid 
excessively long expressions it is better to expand the results for $\xi < 1$. In this limit 
the three roots are:
\begin{eqnarray}
&& \mu_{0} =\frac{213}{4} + 3 \sqrt{14} + {\mathcal O}(\xi), 
\nonumber\\
&& \mu_{\pm} = \biggl(\frac{63}{8} + \frac{9 \sqrt{7}}{\sqrt{2}}\biggr) \pm 6 i\, \frac{2^{1/4}}{7^{1/4} \, \sqrt{\xi}} + {\mathcal O}(\sqrt{\xi}).
\label{FFR4}
\end{eqnarray}
Since all the three roots have positive real parts the solution is not stable.

In summary,  two different anisotropic fixed points 
characterized by different magnetic susceptibilities have been analyzed. The results obtained in the second part of this section 
show, as anticipated, that the existence of magnetic hairs in a given solution does not guarantee the stability of the corresponding 
autonomous system. Thus, the existence of magnetic hairs is not sufficient to warrant the stability of the solution but it is nonetheless necessary 
since the electric hairs are dissipated faster by the protoinflationary dynamics. 

\renewcommand{\theequation}{6.\arabic{equation}}
\setcounter{equation}{0}
\section{Concluding remarks}
\label{sec6}

There is no compelling evidence supporting the physical occurrence of a phase of anisotropic inflationary expansion but it is interesting to scrutinize the validity of the no-hair conjectures by delicately improving the conventional inflationary scenarios. Along this line, new classes of anisotropically inflating solutions have been derived in the context of a generalized gauge action where the electric and magnetic susceptibilities are not bound to coincide. 

The exact solutions derived here have the common feature that the inflationary value of the shear parameter 
depends on the presence of gauge hairs. Since  inflating backgrounds cannot be eternal in the past, the 
criteria to discriminate between physical and unphysical solutions must be connected to the nature 
of the protoinflationary epoch. Magnetic and electric hairs are not physically equivalent since they have different interactions with the protoinflationary plasma: the electric hairs are dissipated much faster than the magnetic ones 
so that the relevant initial conditions will be magnetic rather than electric.  
The existence of magnetic hairs in the solution is a necessary requirement for the survival 
of the primordial shear across the protoinflationary transition but it is not sufficient 
to warrant the stability of the corresponding solution: different classes of magnetic solutions 
may not be stable when perturbed around their corresponding anisotropic 
fixed point. 

The laws of relativistic 
gravitation imply that in conventional quasi-de Sitter backgrounds any finite portion of the Universe gradually 
loses memory of initially imposed anisotropies or inhomogeneities. The present analysis suggests 
that when the initial shear is due caused by gauge hairs, not only gravity but also more standard 
sources of dissipation should be included in a generalized formulation of the no-hair conjecture encompassing all 
the relevant length-scales of the problem. It is our opinion that more work along these directions is desirable.

\section*{Acknowledgements}

It is a pleasure to thank J. Jerdelet and S. A. Rohr of the CERN scientific information service for their kind assistance.

\end{document}